\documentclass[lettersize,journal]{IEEEtran}
\usepackage{amsmath,amsfonts}
\usepackage{algorithmic}
\usepackage{algorithm}
\usepackage{array}
\usepackage[caption=false,font=normalsize,labelfont=sf,textfont=sf]{subfig}
\usepackage{textcomp}
\usepackage{stfloats}
\usepackage{url}
\usepackage{verbatim}
\usepackage{graphicx}
\usepackage{cite}

\DeclareUnicodeCharacter{2212}{-}

\usepackage{multirow}
\usepackage{makecell}
\usepackage{multicol,caption}

\usepackage{xcolor}

\usepackage{mathtools}
\newcommand{\norm}[1]{\| #1 \|}
\newcommand{\bignorm}[1]{\Bigl \| #1 \Bigr \| }

\usepackage{hyperref}
\usepackage{url}

\usepackage{soul}
\usepackage{makecell}

\DeclareUnicodeCharacter{2212}{-}
\DeclareUnicodeCharacter{2026}{-}

\begin{document}



\title{Ground Reaction Force Estimation \\ via Time-aware Knowledge Distillation}



\author{Eun~Som~Jeon,
        Sinjini Mitra,
        Jisoo Lee,
        Omik M. Save,
        Ankita Shukla, \\
        Hyunglae Lee,
        and~Pavan~Turaga

\thanks{E. Jeon is with Department of Computer Science and Engineering, Seoul National University of Science and Technology, Seoul, 01811, Republic of Korea, email: (ejeon6@seoultech.ac.kr).}

\thanks{S. Mitra, J. Lee, and P. Turaga are with Geometric Media Lab, School of Arts, Media and Engineering and School of Electrical, Computer and Energy Engineering, Arizona State University, Tempe, AZ 85281 USA, email: (smitra16@asu.edu; jlee815@asu.edu; pturaga@asu.edu).
}

\thanks{Omik M. Save and H. Lee are with School for Engineering of Matter, Transport and Energy, Arizona State University, AZ 85287 USA, email:
(osave@asu.edu; hyunglae.lee@asu.edu).}

\thanks{A. Shukla is with Department of Computer Science and Engineering, University of Nevada, Reno, NV 89557, USA, email: (ankitas@unr.edu).}
}

\markboth{Journal of \LaTeX\ Class Files,~Vol.~14, No.~8, August~2021}%
{Shell \MakeLowercase{\textit{et al.}}: A Sample Article Using IEEEtran.cls for IEEE Journals}


\maketitle

\begin{abstract}

Human gait analysis with wearable sensors has been widely used in various applications, such as daily life healthcare, rehabilitation, physical therapy, and clinical diagnostics and monitoring. In particular, ground reaction force (GRF) provides critical information about how the body interacts with the ground during locomotion. Although instrumented treadmills have been widely used as the gold standard for measuring GRF during walking, their lack of portability and high cost make them impractical for many applications. As an alternative, low-cost, portable, wearable insole sensors have been utilized to measure GRF; however, these sensors are susceptible to noise and disturbance and are less accurate than treadmill measurements. Deep learning has shown potential in addressing these issues, but such methods are computationally expensive and often require extensive computing resources, limiting their feasibility for real-time and portable systems. To address these challenges, we propose a Time-aware Knowledge Distillation framework for GRF estimation from insole sensor data. This framework leverages similarity and temporal features within a mini-batch during the knowledge distillation process, effectively capturing the complementary relationships between features and the sequential properties of the target and input data. The performance of the lightweight models distilled through this framework was evaluated by comparing GRF estimations from insole sensor data against measurements from an instrumented treadmill. Various teacher-student model architectures and learning strategies were evaluated across multiple performance metrics using data collected at different walking speeds. Empirical results demonstrated that Time-aware Knowledge Distillation outperforms current baselines in GRF estimation from wearable sensor data. Moreover, our method  significantly reduces the number of training parameters needed for GRF estimation, offering a data- and resource-efficient solution for human gait analysis while achieving excellent accuracy and model reliability.

\end{abstract}


\section{Introduction}
\IEEEPARstart{T}{he} study of human internet of things with wearables has been receiving increasing attention for various applications such as daily life analysis for healthcare, physical activity, and smart homes \cite{chen2019bring, cai2022mhealth, dang2020sensor, abdelhady2021design, zhao2024wearable, yang2018smart}. Particularly, analysis of walking and movement has been utilized in disease diagnosis, such as hemiplegia \cite{brunelli2020early} and Parkinson’s
disease \cite{li2021rehabilitation}, both of which make it difficult to maintain balance of movement, leading to instability and irregularity in gait pattern.
Further, for injury prevention and medical therapy, gait signals based on wearables are used to monitor the patient's intention and recovery progress, which can provide information and aid in controlling exoskeleton robots for rehabilitation\cite{ling2022ae}.

For activity monitoring and considerations of mobility, gait analysis with ground reaction force (GRF) sensors has been widely used \cite{xiang2024rethinking, chakraborty2022musculoskeletal, buurke2023comparison}, which is effective and applicable with deep learning based methods \cite{buurke2023comparison, an2023artificial}.
GRF represents the force exerted by the ground on the foot during contact \cite{Key2010}.
Since these forces significantly influence motion, accurate GRF estimation is vital for predicting human lower-body joint movements during dynamic activities \cite{Sakamoto2023}. Additionally, the magnitude and time-based distribution of GRF peaks during walking can play a crucial role in determining lower-limb joint loads \cite{Jiang2020}.
In general, an instrumented treadmill is suitable for measuring GRF signals during walking and running \cite{kluitenberg2012comparison}. However, there are several critical limitations in leveraging GRF signals with treadmill in real-world and real-time system. Since traditional treadmills have challenges related to portability and cost, it is inaccessible to the general public for daily-life use to monitor their health \cite{hong2017developing}. Also, using diverse sensors simultaneously in inference time increases the cost of system implementation \cite{shi2023robust}.

To overcome these issues, recent studies have leveraged in-shoe systems to estimate GRF \cite{burns2019validation, lee2019functional}.
Loadsol (Novel GmbH, Germany), a wireless insole-based plantar load measurement system achieved similar measurement performance to an instrumented treadmill's force plate \cite{burns2019validation}. Similarly, the Tactilus High-Performance V-Series (SensorProd Inc., USA) insole boasts a high density of over 100 sensing points.
However, sensors are impacted by the size of shoes, and tend to drift when subjected to a constant force over a short period, which creates significant errors \cite{chen2024center, Gehlhar2022}. Also, the low quality of sensing units hinders analysis performance \cite{burns2019validation}.
Usually, the errors can be observed during the initial and final stages of the stance phase \cite{dyer2011instrumented}. Additionally, because of system instability, personal variations, and extrinsic gait variability, erroneous signals can be recorded \cite{masani2002variability}.


To alleviate the noisy signal problems and to design an improved system, deep learning based models, such as autoencoders \cite{Hinton2006Dim}, have been utilized.
Generally, adding more network layers improves performance; however, this directly increases the number of trainable parameters in the model, which consequently causes a burden on computing power and storage resources for implementation \cite{huang2017densely, fawaz2019deep, khan2020survey,gil2020improving}.
The demand for real-time inference has recently increased in wearable sensor data analysis, which highlights the importance of developing lightweight models that can process data faster with fewer resources while maintaining the performance comparable to that of larger models \cite{saad2024employing, dargazany2018wearabledl,ni2024survey}.
Additionally, in estimation tasks, conventional deep learning models are typically implemented in a uni-modal manner \cite{Li2022CNN, Lipton2015RNN}, where the input and target data formats and statistical characteristics are the same or highly similar. However, when video data with rich sequential and spatial information is used as input while time-series data, which lacks spatial information, serves as the target, traditional deep learning models tend to show performance degradation.

We address these challenges through two primary objectives: 
1) We propose a framework which aims to estimate GRF from insole sensors effectively while generating a lightweight model. 2) Our framework adopts a multi-modal approach to estimate time-series data from sequential images, video, eliminating the need for treadmill data during testing.
We adopt a knowledge distillation (KD) to estimate GRF, which is one of the promising methods for creating a lightweight model.
KD has been used to generate a smaller model (student) by leveraging the learned knowledge of a larger model (teacher) \cite{hinton2015distilling}. KD methods have shown outstanding performance in the analysis of wearable sensor data \cite{jeon2022kd, gou2021knowledge}. 
In distillation, the information that is delivered for learning plays a key role in creating a better model.

\begin{figure*}[htb!]
\includegraphics[scale=0.43] {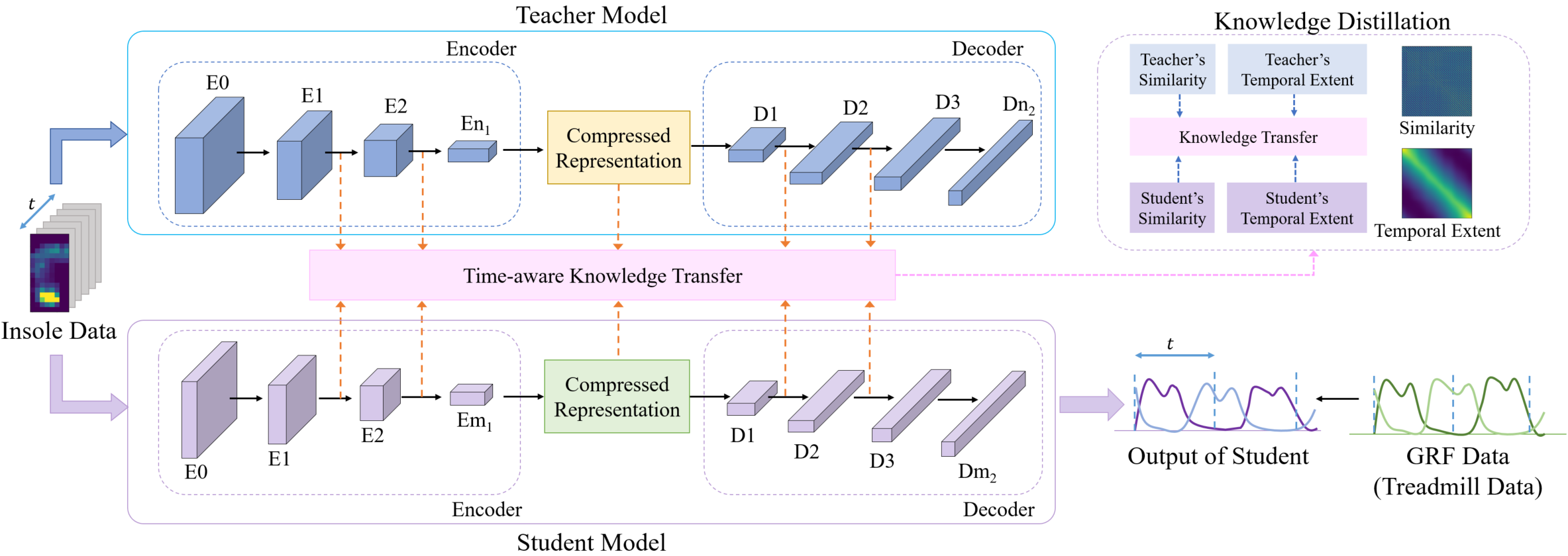} 
\centering
\caption{Overview of Time-aware Knowledge Distillation, TaKD.}
\label{figure:overview}
\end{figure*}

An overview of the proposed approach, Time-aware Knowledge
Distillation (TaKD), is depicted in Figure \ref{figure:overview}.
To elaborate, firstly, a large model is trained by minimizing the distance between generated outputs and GRF data, using insole video data as an input that consists of sequential frames of GRF.
We then train a student with supervision of GRF data and intermediate feature matching. Similarity maps are utilized to extract implicit relationships of intermediate features within a mini-batch.
When there is a large knowledge gap between the input and target data, utilizing similar characteristics between them aids in improving performance \cite{day2017survey, liu2022improved}.
Based on this insight and the fact that the common characteristic between input and target data is sequential knowledge, temporal extent maps are used in the knowledge transfer process, which provides useful information to the student. Finally, a compact model is distilled, which outperforms a model having the same capacity and trained from scratch.

For experiments and investigations from various perspectives, we develop a system architecture to collect the data from an instrumented treadmill and insole sensor, which can collect data more effectively tailored
to satisfy our specific requirements. With the collected data, we are able to analyze our approach with diverse data formats, such as different lengths of window size and number of data.

The contributions of this paper are as follows:
\begin{itemize}
\item We propose a new framework based on knowledge distillation that transfers similarity and temporal extent features to the student for improved GRF estimation, considering sequential and crossmodal characteristics.
\item We develop a system architecture to meet requirements of GRF estimation using insole sensor data, which allows us to explore the effectiveness of our approach in various aspects under different length of window size and number of data.
\item We explore diverse strategies and show empirical results demonstrating the strength of our approach with various metrics and teacher-student combinations on insole sensor data for GRF estimation.
\end{itemize}

The rest of the paper is organized as follows. In section \ref{sec:background}, we provide backgrounds including systems and methods of deep networks. In section \ref{sec:proposed_method}, we introduce our system architecture, data preparation, and the proposed method, a new framework in KD. In section \ref{sec:experiments} and \ref{sec:ablations}, we show our experimental results and analysis. In section \ref{sec:discussion} and \ref{sec:conclusion}, we discuss our findings and conclusions.



\section{Background} \label{sec:background}
\subsection{Deep Networks for Estimation}

There have been attempts to use multi-channel sensor data with deep learning methods for specific tasks \cite{Zhang2023inception, Tian2019fusion}, but most of these efforts focused on simple image classification or action classification tasks. Although one study estimated the Center of Pressure (CoP) using video \cite{Chen2024cop}, the value of CoP was inferred from each frame without considering the temporal information of the video.
In \cite{Mänttäri2021comparison}, 3D CNNs and C-LSTMs are compared in terms of temporal modeling abilities to examine how meaningful features can be extracted from videos, but did not attempt to enhance processing speed for real-time processing. Other studies have aimed to reduce computational cost in sensor data processing using knowledge distillation methods \cite{Mardanpour2023activity, Jeong2022depth}.
A related study \cite{Ling2022step} estimated step length from insole pressure images, utilizing multi-source data that considered the time domain while also aiming to reduce processing time.
Our study explores various CNN-based models that leverage temporal information to estimate GRF from videos composed of insole images, while also using the Knowledge Distillation (KD) method to reduce computation time for real-time processing.

\subsection{Deep Network Training Method} 
Autoencoder (AE), Variational Autoencoder (VAE) \cite{kingma2013auto,fabius2015variational}, and Wasserstein AutoEncoders (WAE) \cite{tolstikhin2018wasserstein} have been widely used for learning representations and solving estimation problems. AE are one of the most common architectures used in representation learning where high dimensional data is embedded to a lower dimension (latent space, $\mathcal{Z}$). An AE learns two functions - an encoding function $(f(\cdot))$ that transforms the input data to a point in $\mathcal{Z}$ and a decoding function $(g(\cdot))$ that reconstructs the input point from $\mathcal{Z}$. The standard auto-encoding objective is to reduce the discrepancy between the original input and the reconstructed output as follows: 
\begin{equation}
    \mathcal{L}_{AE}  = \mathcal{L}_{MSE}(x, g(h_z)),
\end{equation}
where $x$ is the input data and $h_z = f(x)$. In the case of VAEs, this objective is modified to include the variational inference -  the input is mapped to a probability distribution in the latent space instead of a single point. The objective then becomes to not only match the reconstructed output to the original input but to also match the latent distribution to an assumed prior (e.g., gaussian prior with mean $\mu$ and variation $\sigma$). The VAE loss term is defined as follows:
\begin{equation}
    \mathcal{L}_{VAE}  = \mathcal{L}_{MSE}(x, g(h_z)) + \mathcal{L}_{KL}(x, g(h_z)),
\end{equation}
where the first loss term ensures the reconstruction objective and the second term $\mathcal{L}_{KL}$ is KL divergence loss that ensures the distribution matching objective. 
WAEs utilize Wasserstein distance between the model and target distribution, which provides a regularizer to encourage the encoded training distribution to match the prior. The WAE implementation follows the WAE-GAN pipeline where a discriminator is trained in conjunction with the encoder-decoder pair. 
The WAE trainig procedure consists of two steps. In step 1, the encoder $\mathcal{F}$, with encoding function $f(\cdot)$, and decoder $\mathcal{G}$ with a decoding function $g(\cdot)$, are frozen and the latent space is used to obtain  a latent code $z \sim \mathcal{Z}$. $z$ is considered ``fake" since it is not the original input data and in this stage the discriminator, $\chi$ is trained to identify real vs fake latent variables. To this end, the original input data ($x$) is embedded to $\mathcal{Z}$ using the encoder and then passed to the discriminator. The discriminator outputs either 1's or 0's depending on its assessment of whether the input is real vs fake. For $h_z = f(x)$ the discriminator would predict all 1's ideally and hence the discriminator loss is calculated by comparing $\chi(h_z)$ to a sequence of all 1's. 
\begin{equation}
    \mathcal{L}_{\chi} = \frac{1}{N} \sum_i^N \log_e \left(1 - \chi(h_{z_i}) \right).
\end{equation}

In stage 2, $\chi$ is frozen while the encoder-decoder pair are trained on the real data. We input $x$ to the encoder $f$ and then obtain a reconstruction $\hat{x} = g(f(x))$. The reconstruction loss between the input $x$ and $\hat{x}$ is computed as follows:
\begin{equation}
    \mathcal{L}_{recon} = \mathcal{L}_{MSE}(\hat{x}, x).
\end{equation} 
This two step training process is repeated for the entire dataloader. Step 1 ensures that $\chi$ learns to differentiate between real and fake generated data while step 2 pushes the reconstruction performance of the decoder to improve. Thus, $\mathcal{F}, \mathcal{G}$ and $\chi$ are all engaged in an adversarial game where each network tries to outperform the other which results in an overall improvement in generation quality.

\subsection{Knowledge Distillation}
Knowledge distillation is one of the promising methods to generate a compact/small model from a larger model by leveraging the learned knowledge, first introduced by Buciluǎ \emph{et al.} \cite{bucilua2006model}. This was explored and developed more by Hinton \emph{et al.} \cite{hinton2015distilling}. 
Using soft labels, the knowledge is transferred from a particular teacher model to a chosen student model. Soft labels are obtained with temperature hyperparameters and provide superior supervision in matching as well as alleviating overfitting.
The loss function of traditional KD is:
\begin{equation}
    \mathcal{L_{KD}} = \alpha\mathcal{L_{C}} + (1- \alpha) \mathcal{L_{D}},
\end{equation}
where, $\mathcal{L_{C}}$ is cross entropy loss, $\mathcal{L_{D}}$ is KD loss, and $\alpha$ is a hyperparameter to balance two loss terms.
The cross-entropy loss to minimize the difference between output of a student network and the ground-truth is:
\begin{equation}
    \mathcal{L_{C}} = \mathcal{H}(\rho(l_{S}), l_g),
\end{equation}
where, $\rho(\cdot)$ is a softmax function, $\mathcal{H(\cdot)}$ is a cross entropy loss function, $l_S$ is the logits of a student, and $l_g$ is a ground truth label. The difference between the student and teacher are mitigated by KD loss:
\begin{equation}\label{eq3}
    \mathcal{L_{D}} = \tau^{2}\mathcal{L}_{KL}(r_{T}, r_{S}),
\end{equation}
where, $\tau$ is a hyperparameter; $\tau > 1$, $r_T = \rho(l_T/\tau)$ and $r_S = \rho(l_S/\tau)$ are softened outputs of a teacher and student, respectively, and $l_T$ is the logits of a teacher.

To transfer sufficient knowledge and to provide stronger supervision to train the student, features from intermediate layers of a network have often been additionally utilized \cite{gou2021knowledge, zagoruyko2016paying, tung2019similarity}. 
Activation-based attention transfer (AT) \cite{zagoruyko2016paying} is widely used, which uses an attention mapping function to calculate statistics across the channel dimension. To utilize relations of features, similarity-preserving knowledge distillation (SP) \cite{tung2019similarity} was introduced, which matches the similarity within a mini-batch of samples between a teacher and a student.
For similarity map, the dimensional size is determined by the size of mini-batch. Expanding on this concept, the similarity map is computed as follows:
\begin{equation}\label{eq_sp}
 M = F \cdot F^{\top}; F \in \mathbb{R}^{b \times chw},
\end{equation}
where $M \in \mathbb{R}^{b \times b}$ is the similarity map, $F$ is reshaped features from an intermediate layer, $b$ is the size of a mini-batch, and $c$, $h$, and $w$ are the number of channels, height, and width of the output, respectively.
These feature based transfer methods are popularly used; however, these are commonly for matching knowledge in a uni-modal manner when features have similar statistical characteristics.
With this insight, we devise a framework considering temporal extent features in distillation, which is one of crucial keys for handling the features of sequential knowledge and multi-modal manner.

\section{System Architecture and Proposed Method} \label{sec:proposed_method}
The main objective of this study is to introduce a framework which generates a lightweight model as well as effectively estimates GRF from an arrayed low-cost insole in real-world walking conditions. To verify the feasibility of our proposed method, we designed hardware systems for data collection with an instrumented split-belt treadmill and insole wearable sensors, simultaneously.
We introduce the system architectures and the proposed method using deep learning algorithm to generate a lightweight model as follows.

\subsection{Hardware Setup} \label{harset}

\begin{figure}[htb!]
\includegraphics[scale=0.335] {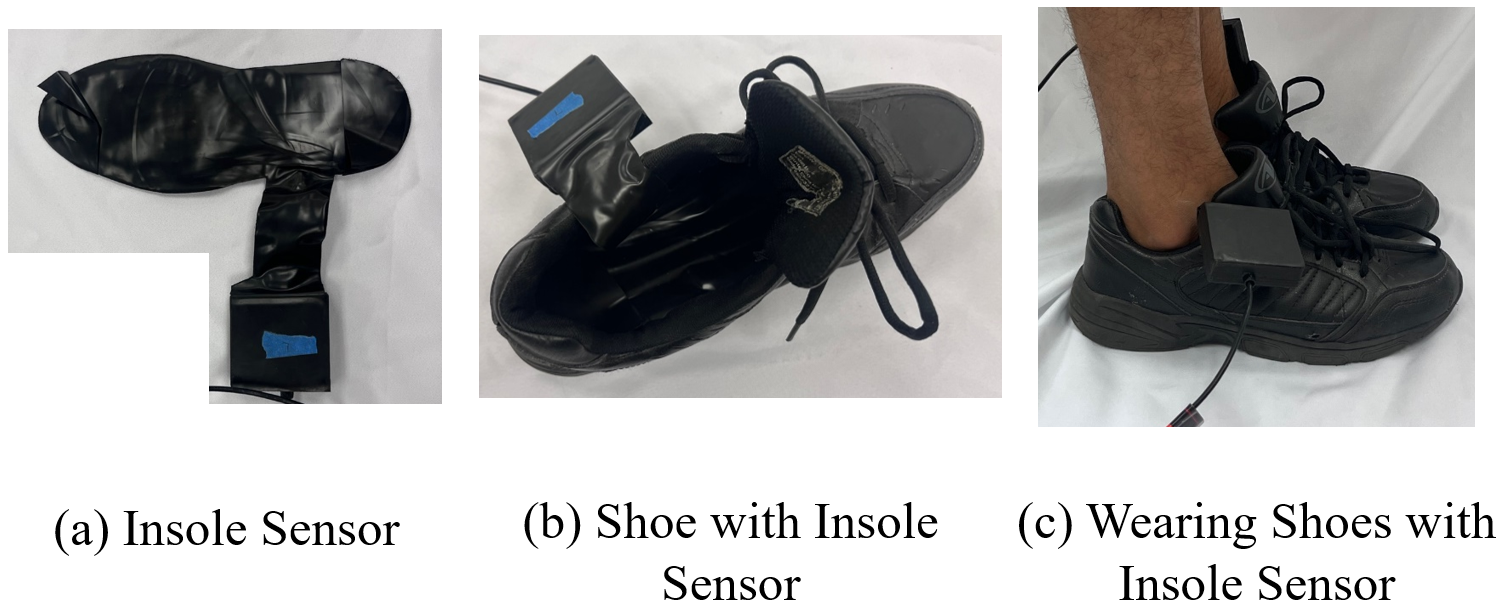} 
\centering
\caption{Examples for insole sensors and shoes.}
\label{figure:insole_sensor}
\end{figure}

In Figure \ref{figure:insole_sensor}, we present the sensor used in this study, the Tactilus V-Series High Performance by SensorProd Inc. (USA), operating in USB mode.
It features 128 piezo-resistive sensors arranged in a 16$\times$8 matrix, with overall dimensions measuring 254.0$\times$92.1 $mm$. The insole sensor is inserted into a shoe, enabling a person to wear it with ease.

\begin{figure}[htb!]
\includegraphics[scale=0.44] {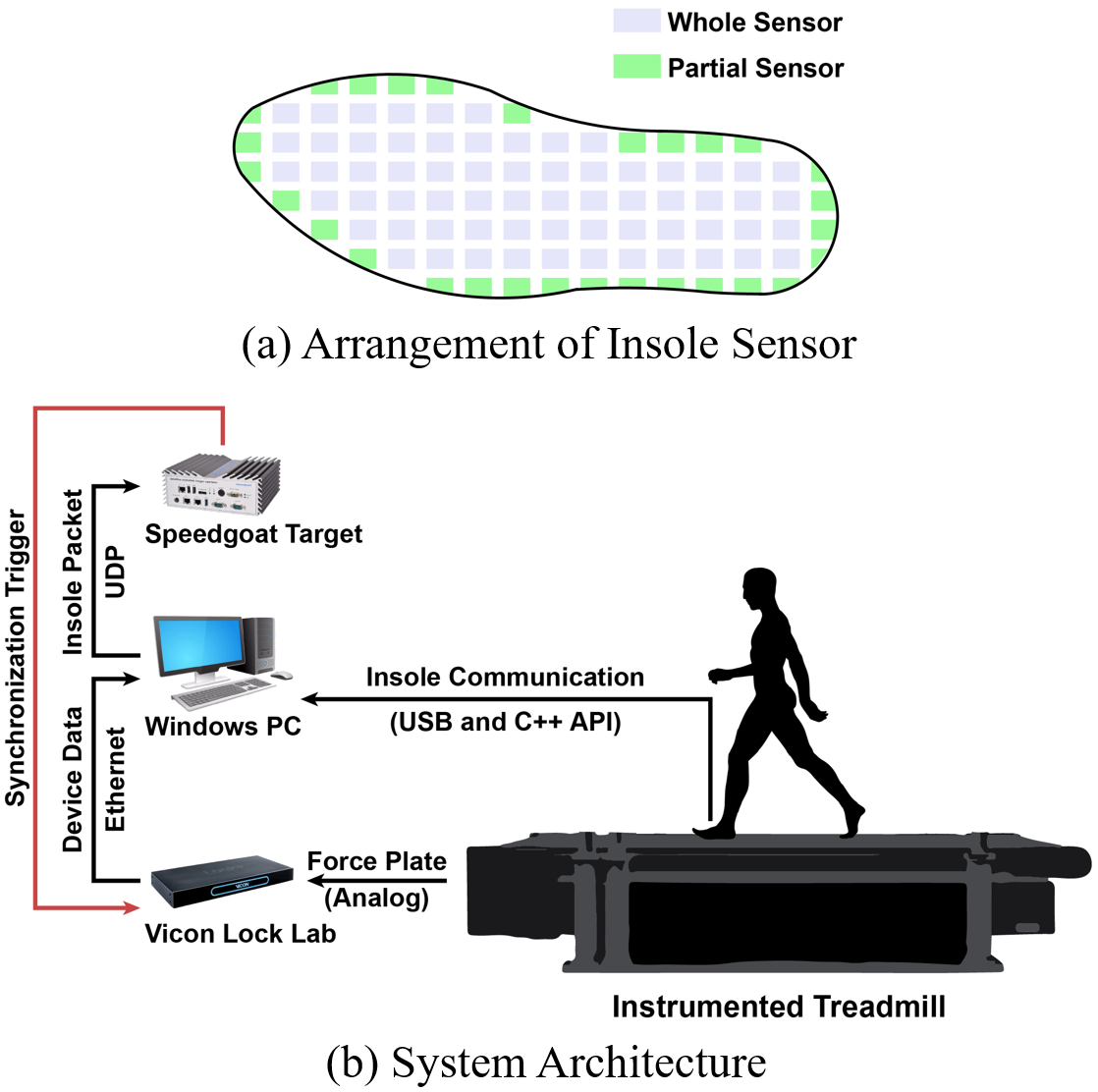} 
\centering
\caption{Arrangement of insole sensor and system architecture for data collection.}
\label{figure:system_architecture}
\end{figure}

Due to the insole's shape, some sensors were either partially present or absent, as depicted in Figure \ref{figure:system_architecture}(a). Each complete sensor could measure pressures up to 206.8 $kPa$ with an accuracy of $\pm$10\%, repeatability of $\pm$2\%, hysteresis of $\pm$5\%, and non-linearity of $\pm$1.5\%. The sensor is interfaced with a Windows computer through a manufacturer-provided API in the form of a precompiled C++ library.

Real-time data collection is facilitated by a multi-threaded C++ program. One thread updated the latest insole frame, while another dedicated thread sent the data to a Speedgoat Baseline Target machine (Speedgoat GmbH, Switzerland) for recording via the User Datagram Protocol (UDP) communication protocol. This simultaneous update minimized sensor delay, allowing refresh rates of up to 200 $Hz$.

Walking experiments were conducted on an instrumented split-belt treadmill (Bertec Inc., USA), with data from embedded force plates recorded, as shown in Figure \ref{figure:system_architecture}(b). Information such as GRF is stored using the Vicon ecosystem (Vicon Motion Systems Ltd., UK) at a rate of 2000 $Hz$. Synchronization is achieved by having the Speedgoat system trigger the Vicon system to initiate data collection remotely. 

\subsection{Data Collection and Preparation}

We recruited eight healthy participants (6M and 2F; age: 29$\pm$5 years; height: 174$\pm$8 $cm$; weight: 64$\pm$6 $kg$) to perform walking tasks at specified treadmill speeds. Due to the insole size restriction, the participant recruitment was constrained to those who wore US 9.5M shoes. The subjects received detailed information about the experiment protocol and their written consent was obtained in compliance with the regulations of Arizona State University's Institutional Review Board (STUDY00014244).

The walking study consisted of 10-minute trials where subjects walked on the treadmill at fixed speed and 0$^\circ$ inclination for the entire duration. The walking speeds investigated in this study are: slow walking (SW) at 0.88 $m/s$, regular walking (RW) at 1 $m/s$, brisk walking (BW) at 1.25 $m/s$, and fast walking (FW) at 1.5 $m/s$. A 10-second buffer was added at the beginning of each walking trial to allow participants to acclimatize to the treadmill speed and this period was not included in data analysis. The four walking conditions were randomized for each participant to prevent any ordering effects of the walking speed changes on the recorded biomechanics. Furthermore, to prevent subject fatigue and insole sensor saturation due to prolonged walking, a 10-minute mandatory break was included between walking trials.

The GRF data obtained from the force plates were downsampled to 200 Hz to match the sampling frequency of insoles and filtered using a zero-lag 2$^{nd}$ order, 10 $Hz$ cut-off, low pass Butterworth filter. The insole data, represented as 16$\times$8$\times n$, where $n$ is the time frame, was spatially filtered using a fraction matrix to account for the effective area of each sensing pixel. Pixels fully within the insole were assigned a value of 1, while boundary pixels were assigned values of 0.33 or 0.67 based on a visual inspection comparing their sizes to a full pixel. Pixels outside the insole boundary or missing due to the insole shape were assigned a value of 0. The insole pressure matrix was then multiplied element-wise by the fraction matrix to obtain the spatially filtered pressure matrix. This approach aimed to mitigate the effects of erroneous load concentrations near the insole boundaries, which were caused by the fabric-like structure of the insole that was susceptible to warping while wearing.

\begin{figure}[htb!]
\includegraphics[scale=0.43] {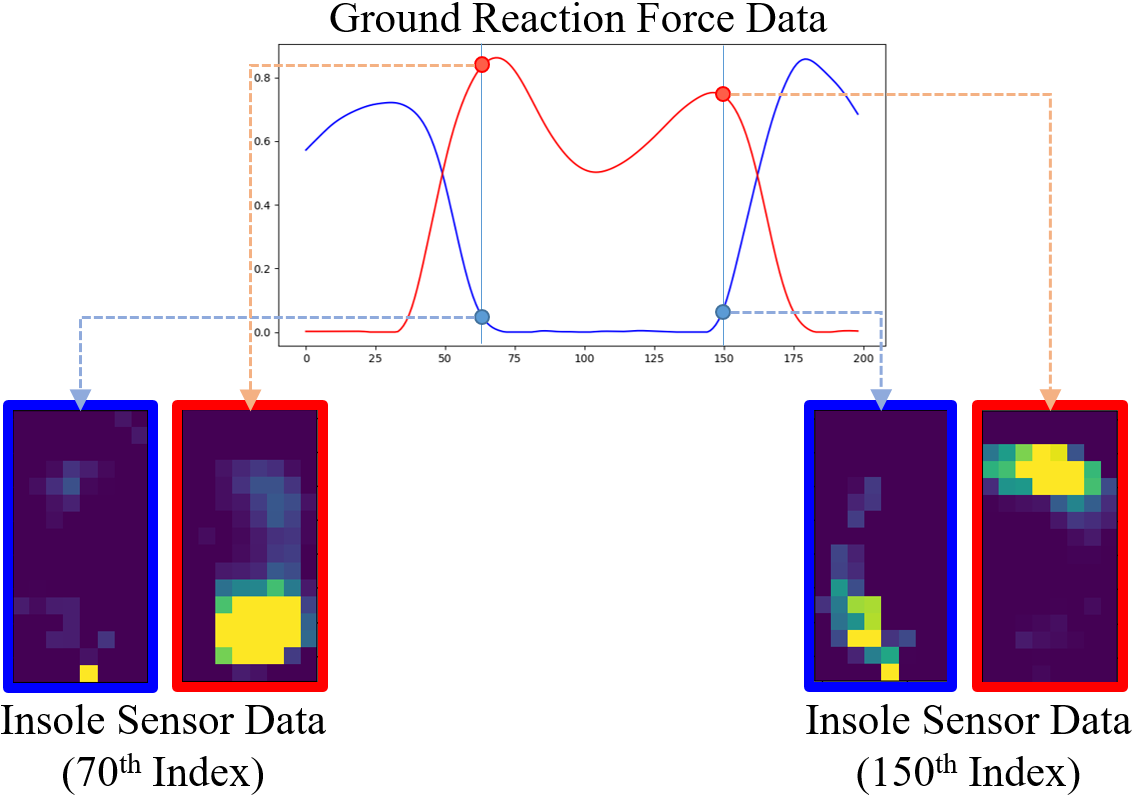} 
\centering
\caption{An example of ground reaction force (GRF) data and its corresponding insole sensor data is presented. In the GRF graph, x and y axes denote time steps and GRF magnitude, respectively. Blue and red lines in the graph correspond to the GRF data for the left and right foot, respectively. The blue and red boxes display the insole sensor data associated with the corresponding GRF data.}
\label{figure:grf_insole}
\end{figure}

In Figure \ref{figure:grf_insole}, an example of ground reaction force data and its corresponding insole sensor data is described.

\subsection{Proposed Method}
\subsubsection{Spatiotemporal Networks}
Using the insole sensor data (ground reaction force) as input, we adopt encoder-decoder networks to generate estimations of the time-series data as ground reaction force.
Since data from insole sensor is collected sequentially, the input data is formatted as a video, which aids in acquiring abundant information about data changes over time.

To encode information from the insole sensor data, we utilize several spatiotemporal convolutional variants, which can preserve temporal information and
propagate the knowledge through the layers of the network.
To construct various architectural designed networks and combinations in KD, we use 3D ConvNets (C3D) \cite{tran2015learning}, Inflated 3D ConvNets (I3D) \cite{carreira2017quo} and (2+1)D \cite{tran2018closer}, which are popularly used to extract feature considering spatiotemporal knowledge. The filters in these network perform convolution computations in the temporal and spatial dimensions.
C3D consists of employing 3D convolutions only. Leveraging the benefit of 2D CNNs, I3D consists of two stream configurations by adopting a 2D CNN architecture, and inflating filters and pooling kernels. (2+1)D factorizes 3D convolution into two successive operations including a 2D spatial convolution and a 1D temporal convolution.


For the decoder, to match the dimension of the encoded representations, 1D CNNs are utilized, which also helps in matching the dimension of the time-series ground truth data.
With these encoder-decoder networks, different types of models can be designed. 
We explain more details about the models that we use for experiments in section \ref{network_architecture}.

\subsubsection{Knowledge Distillation}
To generate a lightweight model, we adopt knowledge distillation, which is a promising solution to create a compact model. We utilize output of decoder, predicted result, and features from intermediate layers for better supervision and to mimic a teacher.

Firstly, we utilize mean squared error loss to minimize the difference between the output of decoder $h_d$ and ground truth $h_g$ that is the corresponding data from treadmill, as follows:
\begin{equation}
    \mathcal{L}_{gt} = \mathcal{L}_{MSE}(h_d, h_g).
\end{equation}

To provide sufficient knowledge from a teacher to a student, we use feature-based knowledge distillation. We utilize similarity within a mini-batch \cite{tung2019similarity}, which is explained in equation \eqref{eq_sp}. 
Specifically, the similarity map $G \in \mathbb{R}^{b \times b}$ from 3D CNNs is computed as follows:
\begin{equation}\label{eq_bs}
 G = F \cdot F^{\top}; F \in \mathbb{R}^{b \times cthw},
\end{equation}
where $t$ is the temporal extent for the feature.
For 1D CNNs, $G$ is obtained from $F \in \mathbb{R}^{b \times ct}$.
Even though the size of intermediate feature pairs for teacher and student can be differ, this method aids in matching the size of similarity maps easily, where the size is determined by the size of a mini-batch. Also, this enables to extract useful relational features between samples.
The loss function that enables the student to mimic teachers with similarity maps is:
\begin{equation}
    \mathcal{L}_{bs} = \frac{1}{b^2|L_{1}|}
    \sum_{(l^T, l^S) \in L} \biggl( \bignorm{ \frac{G^{(l^T)}_{T}}{\norm{G^{(l^T)}_{T}}} - \frac{G^{(l^S)}_{S}}{\norm{G^{(l^S)}_{S}}} }^{2}_{F} \biggr),
\end{equation}
where $G^{(l^T)}_{T}$ and $G^{(l^S)}_{S}$ are similarity map of layer pairs ($l^T$ and $l^S$) from a teacher and a student, respectively, $\norm{\cdot}_F$ is the Frobenius norm, and $L_{1}$ accumulates the layer pairs.

Even if input as a video and target
as a time series data have different types of representations, they have common properties such as possessing consecutive information.
With the insight, to utilize temporal properties for knowledge transfer, we use relations representing temporal properties which can be calculated as follows:
\begin{equation}\label{eq_bt}
 P = F \cdot F^{\top}; F \in \mathbb{R}^{t \times bchw},
\end{equation}
where $P \in \mathbb{R}^{t \times t}$ is a temporal extent map. For 1D CNNs, $P$ is computed by $F \in \mathbb{R}^{t \times bc}$.
The loss function to transfer temporal knowledge from a teacher to a student is:
\begin{equation}
    \mathcal{L}_{tp} = \frac{1}{|L_{2}|}
    \sum_{(l^T, l^S) \in L} \frac{1}{t^2_{l^T}} \biggl( \bignorm{ \frac{P^{(l^T)}_{T}}{\norm{P^{(l^T)}_{T}}} - \frac{P^{(l^S)}_{S}}{\norm{P^{(l^S)}_{S}}} }^{2}_{F} \biggr),
\end{equation}
where $P^{(l^T)}_{T}$ and $P^{(l^S)}_{S}$ are maps implying temporal properties of layer pairs ($l^T$ and $l^S$) from a teacher and a student, respectively, and $L_{2}$ collects the layer pairs.

\begin{figure}[htb!]
\includegraphics[scale=0.5] {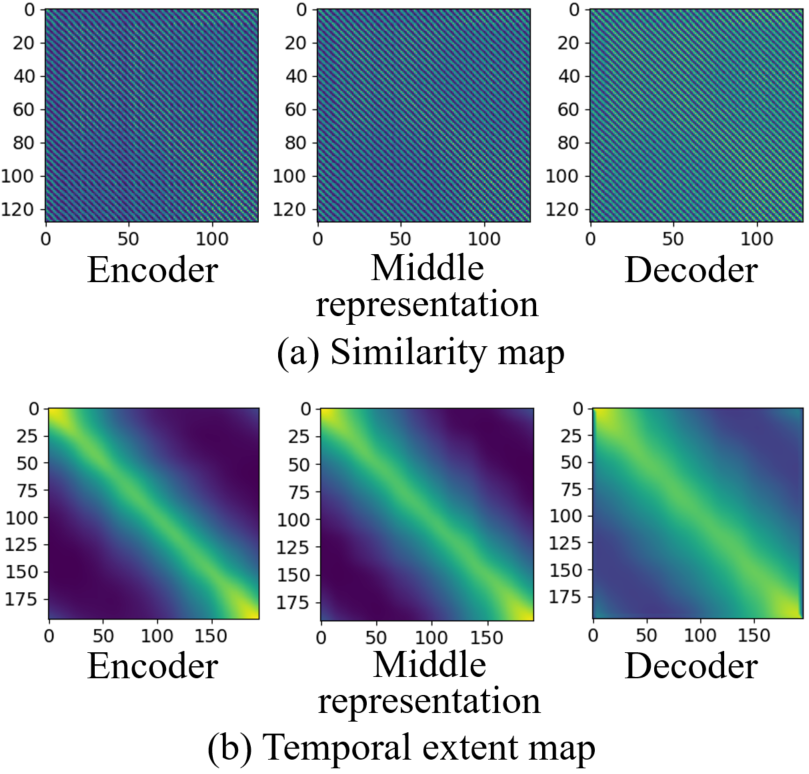} 
\centering
\caption{Illustration of maps representing similarity and temporal properties of intermediate features.}
\label{figure:inter_map}
\end{figure}

As illustrated in Figure \ref{figure:inter_map}, similarity and temporal properties capture different characteristics of intermediate layers. This implies that utilizing both knowledge can provide different aspects of information and complement each other. High similarities for samples within a mini-batch are represented with high values.

In the end, the learning objective of the proposed method can be written as: 
\begin{equation}
    \mathcal{L}_\text{BT} = \mathcal{L}_{gt} + \lambda_1(\mathcal{L}_{{bs}_{m}} +
    \kappa\mathcal{L}_{{tp}_{m}}) + \lambda_2(\mathcal{L}_{{bs}_{i}} +
    \kappa\mathcal{L}_{{tp}_{i}}),
\end{equation}
where, $\lambda_1$ and $\lambda_2$ are hyperparameters to balance between compressed representations (output of the encoder) and intermediate features of encoder and decoder, $\kappa$ is a constant value to control the effects of similarity and temporal extent, $\mathcal{L}_{{bs}_{m}}$ and $\mathcal{L}_{{tp}_{m}}$ are loss values for compressed representations of similarity and temporal extent features, and $\mathcal{L}_{{bs}_{i}}$ and $\mathcal{L}_{{tp}_{i}}$ are loss values for intermediate features, respectively.



\subsubsection{Knowledge Transfer in  Intermediate Layers}
To provide sufficient knowledge to a student, we utilize compressed representations (output of encoder) with both similarity and temporal extent maps.

We use different types of encoders to construct diverse teachers. For a student, C3D encoder is leveraged.
Since the architectural designs of a teacher and student networks are different, the sizes of temporal maps from a teacher and student can differ.
When the sizes of temporal maps from a teacher and student are not the same, bilinear interpolation is applied to the map from student to match the size of the one from teacher.

Not only middle representation, but also intermediate layers of encoder and decoder networks can be used in distillation to improve performance. 
As explored in prior studies \cite{jeong2022lightweight}, high- or mid-level layers provide different effects. When the layers of encoder and decoder near the middle representation (output of encoder) are additionally used with similarity and temporal extent maps in knowledge transfer, we denote this approach as TaKD$\dagger$. When more layers are selected compared to TaKD$\dagger$, we denote this approach as TaKD$\ddagger$. For TaKD$\ddagger$, layers that are far from the middle layer are used for only similarity knowledge transfer. The details for layer selections of TaKD are explained in section \ref{layer_selc}.

\section{Experimental Results and Analysis} \label{sec:experiments}

In this section, experimental datasets and settings are explained. The proposed algorithm is evaluated with different length of window size for samples and various combinations of teacher-student with different strategies.

\subsection{Experimental Settings} \label{sec:settings}

\subsubsection{Data Description}


Commonly, sequential data analysis is affected by the change of window length.
To demonstrate the robustness to change in window lengths, we set different window length settings for the dataset with a sliding window procedure (full-non-overlapping sliding windows).

\begin{table*}[htb!]
\caption{Details of collected dataset. The dataset consists of different speeds recorded for 8 subjects.}
\label{table:Dataset}
\centering
\begin{tabular}{c|c| c c c c c c c c |c c}
\hline
\centering
  & Walking speed & Sbj.01 & Sbj.02 &
Sbj.03 & Sbj.04 & Sbj.05 & Sbj.06 & 
Sbj.07 & Sbj.08 & Sum & \# of subjects\\
\hline
 \multirow{4}{*}{\rotatebox[origin=c]{90}{W200}}  & SW & 279 & 279 & 279 & 279 & 279 & 279 & 279 & 279 & 2232 & 8\\ 
   & RW & 0 & 0 & 279 & 279 & 279 & 279 & 279 & 279 & 1674 & 6\\ 
   & BW & 0 & 279 & 279 & 279 & 279 & 279 & 279 & 279 & 1953 & 7\\ 
   & FW & 279 & 279 & 279 & 0 & 279 & 279 & 279 & 279 & 1953 & 7\\ \hline
 \multirow{4}{*}{\rotatebox[origin=c]{90}{W100}}  & SW & 559 & 559 & 559 & 559 & 559 & 559 & 559 & 559 & 4472 & 8\\ 
   & RW & 0 & 0 & 559 & 559 & 559 & 559 & 559 & 559 & 3354 & 6\\ 
   & BW & 0 & 559 & 559 & 559 & 559 & 559 & 559 & 559 & 3913 & 7\\ 
   & FW & 559 & 559 & 559 & 0 & 559 & 559 & 559 & 559 & 3913 & 7\\
\hline
\end{tabular}
\end{table*}

In most cases, for the data collected, one gait cycle is represented with around 200 window size. To consider this, we set 200 window length as a default size. To evaluate different window length sizes, we set another window size to 100, which is more challenging and potentially shows more performance gaps between methods.
We explain the statistics of datasets in Table \ref{table:Dataset}. SW, RW, BW, and FW denote walking speeds of 0.88, 1.0, 1.25, and 1.5 $m/s$, respectively. The distribution of the dataset is not balanced for each window size. Since additional pre-processing or post-processing is not applied for the data, samples are not smoothened and have perturbations, which creates challenges in estimation.

We first normalized the GRF data by each subject's bodyweight (expressed as \% bodyweight) based on measurements from the treadmill and the insole data (\textit{n}-pixels measuring 0-30 \textit{psi} each). Additionally, the data was further normalized to the range $[0, 1]$ for model training and validation. The normalization range was determined using the absolute difference between the maximum and minimum observed values in the dataset for each variable. Since we analyze data for both feet simultaneously, the size of channels for the treadmill and insole data is 2.

\subsubsection{Network Architectures} \label{network_architecture}


\begin{table*}[htb!]
\centering
\caption{Details of teacher and student network architectures. Brackets denote network architecture type of encoder. * denotes small size of network.}

\begin{center}
\begin{tabular}{c |c |c |c |c |c |c | c}
\hline
\centering
\multirow{2}{*}{Window}& \multirow{2}{*}{Teacher} & \multirow{2}{*}{Student} & \multicolumn{2}{c|}{FLOPs} & \multicolumn{2}{c|}{\# of params} & Compression  \\  \cline{4-7}
& & & Teacher & Student & Teacher & Student & ratio  \\
 \hline 
\multirow{3}{*}{200}& Teacher1 (C3D) & \multirow{3}{*}{C3D$^*$} & 635.39M & \multirow{3}{*}{164.55M} & 1.00M & \multirow{3}{*}{0.25M} & 25.10$\%$ \\ 
& Teacher2 (I3D) &  & 745.90M &  & 1.18M & & 21.35$\%$ \\ 
& Teacher3 ((2+1)D)&  & 1580.11M & & 1.58M & & 15.88$\%$ \\ 
\hline

\multirow{3}{*}{100}& Teacher1 (C3D) & \multirow{3}{*}{C3D$^*$} & 308.66M & \multirow{3}{*}{79.98M} & 1.00M & \multirow{3}{*}{0.25M} & 25.10$\%$ \\ 
& Teacher2 (I3D) &  & 373.51M &  & 1.18M & & 21.35$\%$ \\ 
& Teacher3 ((2+1)D)&  & 790.61M & & 1.58M & & 15.88$\%$ \\ 
\hline

\end{tabular}
\end{center}

\label{table:info_settings}
\end{table*}

To construct different types of teachers and students, we utilize C3D, I3D, and (2+1)D structures as an encoder, which have widely leveraged for video and temporal image analysis \cite{ji20123d, huang2023review}. Let input dimension size is set as
($b, c_i, t_i, h_i, w_i$). Let ($k_1$, $k_2$, $k_3$) is the kernel size $k_1 \times k_2 \times k_3$ for ($t_i$, $h_i$, $w_i$).
For C3D encoder, the network is constructed with $4$ convolutional layers with ReLU. The network referred to the standard 3D convolutional layers. For the first layer, $k_2$ is set as 4 and corresponding stride is 2.
The decoder consists of kernel size 3 of $4$ convolutional layers with ReLU.
%
For I3D encoder, the network is constructed using $8$ convolutional layers with ReLU. The encoder follows the traditional I3D architecture \cite{carreira2017quo}.
For the network that we use, in the first layer, kernel size is set as (5, 4, 3), stride is (1, 2, 1), and padding is (2, 0, 0). For 7th convolutional layer, $k_3$ is set as 2. When $k_1$ is 3 for kernel for convolution, 1 for padding is applied. 
The decoder consists of kernel size $5$ of $4$ convolutional layers with ReLU.
For (2+1)D encoder, the network is constructed using $5$ convolutional layers with ReLU. This network follows the conventional (2+1)D settings. For the first layer, kernel is set as (1, 4, 3) and stride is (1, 2, 1). For second layer, $k_1$ is 5 and corresponding padding is 2. For the forth layer, to consider better encoding spatiotemporal characteristics, (3, 3, 3) kernel and (1, 0, 0) padding are applied. In the final layer, $k_3$ is 2. The decoder structure is the same with the one used for I3D encoder. We construct a student model with small sized networks of C3D encoder and 1D decoder of CNNs. This network design is to construct different architectural combinations for teacher-student models.

With these network types, teacher and student combinations are defined, which is explained in Table \ref{table:info_settings}. Teachers consisting of C3D, I3D, and (2+1)D encoders are denoted as Teacher1, Teacher2, and Teacher3, respectively.
To explain the model complexity, we report floating point operations per second (FLOPs) and the number of trainable parameters with model compression ratio for different window sizes, where FLOPs is a measure of computer performance in computing.
As explained in the table, compared to teachers, a student model possesses much smaller number of parameters and requires much less operations. We choose a student model consisting of C3D encoder since this network has smaller size than the one of other types of encoders.

\subsubsection{Training and Evaluation}
For training models, we use a total of 200 epochs with a batch size of 128.
We utilize the Adam optimizer with the learning rate of 0.01 for C3D and I3D encoders, and 0.001 for (2+1)D encoder utilization. 
Empirically, hyperparameters of loss functions are determined. $\lambda_{1}$ and $\lambda_{2}$ are 0.01 and 0.1, respectively. $\kappa$ is set to 0.1.

To interpret model performance, we report the root mean squared error (RMSE, $\times10^{-2}$), mean absolute error (MAE, $\times10^{-2}$), and Pearson correlation coefficient (represented as $r$, $\times10^{-2}$) between the predicted data from insole sensor and measured data from treadmill, referring to the previous studies \cite{orlin2000plantar}. The correlation coefficient is the ratio between the covariance of two variables and the product of their standard deviations. If the correlation is high, the value is close to 1. We run three times and report the averaged results with standard deviation.
To consider evaluation with generalizability of models on subjects, we utilize leave-one-subject-out metric for evaluation.
We compare our method with conventional KD \cite{hinton2015distilling}, attention transfer (AT) \cite{zagoruyko2016paying}, similarity-preserving knowledge distillation (SP) \cite{tung2019similarity}, DIST \cite{huang2022knowledge}, and semantic calibration for cross-layer knowledge distillation (SemCKD) \cite{semckd}. Considering the applicability of the baselines, middle representation is leveraged in knowledge transfer. For DIST, considering its original protocols, the output of decoder is utilized. $\lambda$ hyperparameters for distillation of DIST and SemCKD are set empirically as 1. Hyperparameters for other baselines are the same with our method.

\subsection{Analysis on Network Architectures}


\begin{table}[htb!]
\centering
\caption{Comparison of predicted ground reaction force results with other methods. \textbf{Bold} and \textcolor{red}{red} denote best and second-best RMSE and MAE for a student, respectively.}

\begin{center}
\begin{tabular}{c |c c c |c c c}

\hline
\multirow{2}{*}{Method} & \multicolumn{3}{c|}{W200} & \multicolumn{3}{c}{W100} \\ \cline{2-7}



& RMSE & MAE & $r$ & RMSE & MAE & $r$ \\ \hline

\multirow{2}{*}{Student} & 6.253 & 4.558 & 98.757 & 6.458 & 4.731 & 94.246 \\
& {\scriptsize$\pm$0.259} & {\scriptsize$\pm$0.162} & {\scriptsize$\pm$0.071} & {\scriptsize$\pm$0.229} &
{\scriptsize$\pm$0.158} & {\scriptsize$\pm$0.390} \\ \hline \hline

\multirow{2}{*}{Teacher1} & 5.978 & 4.376 & 98.740 & 6.321 & 4.648 & 94.552 \\
& {\scriptsize$\pm$0.189} & {\scriptsize$\pm$0.117} & {\scriptsize$\pm$0.073} & {\scriptsize$\pm$0.216} &
{\scriptsize$\pm$0.154} & {\scriptsize$\pm$0.334} \\ \hline

\multirow{2}{*}{KD} & 6.264 & 4.577 & 98.678 & 6.476 & 4.764 & 94.466 \\
& {\scriptsize$\pm$0.248} & {\scriptsize$\pm$0.159} & {\scriptsize$\pm$0.089} & {\scriptsize$\pm$0.209} &
{\scriptsize$\pm$0.145} & {\scriptsize$\pm$0.390} \\

\multirow{2}{*}{AT} & 6.232 & 4.550 & 98.708 & 6.507 & 4.785 & 94.293 \\
& {\scriptsize$\pm$0.236} & {\scriptsize$\pm$0.150} & {\scriptsize$\pm$0.078} & {\scriptsize$\pm$0.235} &
{\scriptsize$\pm$0.165} & {\scriptsize$\pm$0.390} \\


\multirow{2}{*}{SP} & 6.228 & 4.560 & 98.705 & 6.482 & 4.742 & 94.618 \\
& {\scriptsize$\pm$0.233} & {\scriptsize$\pm$0.151} & {\scriptsize$\pm$0.078} & {\scriptsize$\pm$0.237} &
{\scriptsize$\pm$0.165} & {\scriptsize$\pm$0.416} \\

\multirow{2}{*}{KD+SP} & 6.248 & 4.582 & 98.707 & 6.465 & 4.726 & 94.379 \\
& {\scriptsize$\pm$0.237} & {\scriptsize$\pm$0.153} & {\scriptsize$\pm$0.080} & {\scriptsize$\pm$0.238} &
{\scriptsize$\pm$0.168} & {\scriptsize$\pm$0.421} \\

\multirow{2}{*}{DIST} & \textbf{6.106} & \textbf{4.495} & 98.779 & \textbf{6.260} & \textbf{4.627} & 94.879 \\
& {\scriptsize$\pm$0.214} & {\scriptsize$\pm$0.128} & {\scriptsize$\pm$0.075} & {\scriptsize$\pm$0.141} &
{\scriptsize$\pm$0.085} & {\scriptsize$\pm$0.336} \\

\multirow{2}{*}{SemCKD} & 6.221 & 4.524 & 98.631 & 6.400 & 4.719 & 93.749 \\
& {\scriptsize$\pm$0.181} & {\scriptsize$\pm$0.112} & {\scriptsize$\pm$0.070} & {\scriptsize$\pm$0.163} &
{\scriptsize$\pm$0.112} & {\scriptsize$\pm$0.433} \\ \hline

\multirow{2}{*}{TaKD} & 6.220 & 4.546 & 98.682 & 6.375 & \textcolor{red}{4.658} & 94.398 \\
& {\scriptsize$\pm$0.234} & {\scriptsize$\pm$0.152} & {\scriptsize$\pm$0.087} & {\scriptsize$\pm$0.183} &
{\scriptsize$\pm$0.122} & {\scriptsize$\pm$0.414} \\

\multirow{2}{*}{TaKD$\dagger$} & \textcolor{red}{6.213} & \textcolor{red}{4.516} & 98.705 & \textcolor{red}{6.369} & 4.669 & 94.451 \\
& {\scriptsize$\pm$0.252} & {\scriptsize$\pm$0.153} & {\scriptsize$\pm$0.082} & {\scriptsize$\pm$0.194} &
{\scriptsize$\pm$0.130} & {\scriptsize$\pm$0.375} \\

\multirow{2}{*}{TaKD$\ddagger$} & 6.249 & 4.553 & 98.691 & 6.488 & 4.749 & 94.175 \\
& {\scriptsize$\pm$0.248} & {\scriptsize$\pm$0.154} & {\scriptsize$\pm$0.085} & {\scriptsize$\pm$0.217} &
{\scriptsize$\pm$0.146} & {\scriptsize$\pm$0.393} \\ \hline \hline

\multirow{2}{*}{Teacher2} & 5.382 & 3.962 & 99.027 & 5.520 & 4.056 & 96.306 \\
& {\scriptsize$\pm$0.226} & {\scriptsize$\pm$0.163} & {\scriptsize$\pm$0.084} & {\scriptsize$\pm$0.252} &
{\scriptsize$\pm$0.185} & {\scriptsize$\pm$0.280} \\ \hline

\multirow{2}{*}{KD} & 6.293 & 4.604 & 98.694 & 6.430 & 4.723 & 94.369 \\
& {\scriptsize$\pm$0.277} & {\scriptsize$\pm$0.184} & {\scriptsize$\pm$0.081} & {\scriptsize$\pm$0.233} &
{\scriptsize$\pm$0.168} & {\scriptsize$\pm$0.388} \\

\multirow{2}{*}{AT} & 6.268 & 4.543 & 98.652 & 6.460 & 4.751 & 94.475 \\
& {\scriptsize$\pm$0.268} & {\scriptsize$\pm$0.167} & {\scriptsize$\pm$0.102} & {\scriptsize$\pm$0.227} &
{\scriptsize$\pm$0.163} & {\scriptsize$\pm$0.353} \\

\multirow{2}{*}{SP} & 6.297 & 4.600 & 98.668 & 6.460 & 4.723 & 94.369 \\
& {\scriptsize$\pm$0.263} & {\scriptsize$\pm$0.174} & {\scriptsize$\pm$0.087} & {\scriptsize$\pm$0.230} &
{\scriptsize$\pm$0.164} & {\scriptsize$\pm$0.396} \\

\multirow{2}{*}{KD+SP} & 6.304 & 4.615 & 98.674 & 6.442 & 4.738 & 94.341 \\
& {\scriptsize$\pm$0.246} & {\scriptsize$\pm$0.161} & {\scriptsize$\pm$0.078} & {\scriptsize$\pm$0.231} &
{\scriptsize$\pm$0.164} & {\scriptsize$\pm$0.390} \\

\multirow{2}{*}{DIST} & \textbf{6.098} & \textbf{4.406} & 98.789 & 6.398 & 4.698 & 94.517 \\
& {\scriptsize$\pm$0.172} & {\scriptsize$\pm$0.124} & {\scriptsize$\pm$0.052} & {\scriptsize$\pm$0.212} &
{\scriptsize$\pm$0.164} & {\scriptsize$\pm$0.256} \\

\multirow{2}{*}{SemCKD} & \textcolor{red}{6.197} & \textcolor{red}{4.490} & 98.660 & \textcolor{red}{6.374} & 4.722 & 93.723 \\
& {\scriptsize$\pm$0.196} & {\scriptsize$\pm$0.126} & {\scriptsize$\pm$0.071} & {\scriptsize$\pm$0.198} &
{\scriptsize$\pm$0.137} & {\scriptsize$\pm$0.382} \\ \hline

\multirow{2}{*}{TaKD} & 6.232 & 4.547 & 98.697 & 6.457 & 4.744 & 94.349 \\
& {\scriptsize$\pm$0.246} & {\scriptsize$\pm$0.161} & {\scriptsize$\pm$0.078} & {\scriptsize$\pm$0.231} &
{\scriptsize$\pm$0.165} & {\scriptsize$\pm$0.399} \\

\multirow{2}{*}{TaKD$\dagger$} & 6.217 & 4.522 & 98.674 & \textbf{6.367} & \textbf{4.683} & 94.341 \\
& {\scriptsize$\pm$0.237} & {\scriptsize$\pm$0.143} & {\scriptsize$\pm$0.091} & {\scriptsize$\pm$0.195} &
{\scriptsize$\pm$0.129} & {\scriptsize$\pm$0.418} \\

\multirow{2}{*}{TaKD$\ddagger$} & 6.256 & 4.559 & 98.669 & 6.414 & \textcolor{red}{4.696} & 94.214 \\
& {\scriptsize$\pm$0.240} & {\scriptsize$\pm$0.147} & {\scriptsize$\pm$0.088} & {\scriptsize$\pm$0.217} &
{\scriptsize$\pm$0.142} & {\scriptsize$\pm$0.423} \\ \hline \hline

\multirow{2}{*}{Teacher3} & 5.623 & 4.091 & 98.849 & 5.754 & 4.188 & 95.782 \\
& {\scriptsize$\pm$0.252} & {\scriptsize$\pm$0.163} & {\scriptsize$\pm$0.124} & {\scriptsize$\pm$0.306} &
{\scriptsize$\pm$0.203} & {\scriptsize$\pm$0.473} \\ \hline

\multirow{2}{*}{KD} & 6.284 & 4.584 & 98.715 & 6.554 & 4.814 & 94.291 \\
& {\scriptsize$\pm$0.271} & {\scriptsize$\pm$0.176} & {\scriptsize$\pm$0.074} & {\scriptsize$\pm$0.241} &
{\scriptsize$\pm$0.166} & {\scriptsize$\pm$0.373} \\

\multirow{2}{*}{AT} & 6.245 & 4.543 & 98.711 & 6.487 & 4.771 & 94.311 \\
& {\scriptsize$\pm$0.266} & {\scriptsize$\pm$0.168} & {\scriptsize$\pm$0.088} & {\scriptsize$\pm$0.235} &
{\scriptsize$\pm$0.166} & {\scriptsize$\pm$0.390} \\

\multirow{2}{*}{SP} & 6.236 & 4.544 & 98.672 & 6.606 & 4.842 & 94.176 \\
 & {\scriptsize$\pm$0.258} & {\scriptsize$\pm$0.168} & {\scriptsize$\pm$0.092} & {\scriptsize$\pm$0.254} &
{\scriptsize$\pm$0.173} & {\scriptsize$\pm$0.395} \\

\multirow{2}{*}{KD+SP} & 6.221 & \textcolor{red}{4.530} & 98.675 & 6.586 & 4.842 & 94.176 \\
& {\scriptsize$\pm$0.253} & {\scriptsize$\pm$0.163} & {\scriptsize$\pm$0.091} & {\scriptsize$\pm$0.253} &
{\scriptsize$\pm$0.173} & {\scriptsize$\pm$0.396} \\

\multirow{2}{*}{DIST} & 6.245 & 4.537 & 98.682 & 6.525 & 4.748 & 95.084 \\
& {\scriptsize$\pm$0.272} & {\scriptsize$\pm$0.165} & {\scriptsize$\pm$0.106} & {\scriptsize$\pm$0.214} &
{\scriptsize$\pm$0.163} & {\scriptsize$\pm$0.471} \\

\multirow{2}{*}{SemCKD} & 6.315 & 4.574 & 98.548 & 6.483 & 4.778 & 93.583 \\
& {\scriptsize$\pm$0.184} & {\scriptsize$\pm$0.098} & {\scriptsize$\pm$0.083} & {\scriptsize$\pm$0.172} &
{\scriptsize$\pm$0.130} & {\scriptsize$\pm$0.339} \\ \hline

\multirow{2}{*}{TaKD} & \textcolor{red}{6.220} & 4.532 & 98.680 & 6.580 & 4.819 & 94.235 \\
& {\scriptsize$\pm$0.255} & {\scriptsize$\pm$0.164} & {\scriptsize$\pm$0.092} & {\scriptsize$\pm$0.252} &
{\scriptsize$\pm$0.172} & {\scriptsize$\pm$0.389} \\

\multirow{2}{*}{TaKD$\dagger$} & \textbf{6.201} & \textbf{4.528} & 98.732 & \textcolor{red}{6.394} & \textcolor{red}{4.720} & 94.414 \\
& {\scriptsize$\pm$0.257} & {\scriptsize$\pm$0.173} & {\scriptsize$\pm$0.072} & {\scriptsize$\pm$0.189} &
{\scriptsize$\pm$0.123} & {\scriptsize$\pm$0.406} \\

\multirow{2}{*}{TaKD$\ddagger$} & 6.217 & 4.561 & 98.711 & \textbf{6.381} & \textbf{4.705} & 94.372 \\
& {\scriptsize$\pm$0.250} & {\scriptsize$\pm$0.168} & {\scriptsize$\pm$0.073} & {\scriptsize$\pm$0.184} &
{\scriptsize$\pm$0.123} & {\scriptsize$\pm$0.399} \\

\hline

\end{tabular}
\end{center}
\label{table:grf_w100w200}
\end{table}


To analyze the performance on different combinations of teachers and students, we utilize different architectural networks including C3D, I3D, and (2+1)D encoder based teachers (Teacher1, Teacher2, and Teacher3) with AE.
C3D network is used to construct a student, however its capacity is much smaller than teachers, which were more explained in the previous section.
We evaluate the predicted ground reaction force results on window size of 100 and 200 with evaluation metrics of RMSE, MAE, and $r$.
The results of different knowledge distillation methods on network architectures and various window lengths are explained in Table \ref{table:grf_w100w200}.
As explained in the table, TaKD$\dagger$ shows the stable results in most of the cases, showing best or second-best results that are lower RMSE and MAE, and higher $r$ that is over 90$\%$.
For DIST, only when the network structures of teacher and student models are similar, better results of student model can be obtained. Also, a specific condition is required  Furthermore, the methods, such as DIST and SemCKD, require to modify networks to match the feature dimensional sizes for loss computation or utilize hidden layers in KD process, which increases more time consumption and resources.
For window size of 100, in most of the cases, the results of RMSE and MAE from ours and baselines are higher than results of window size of 200. This implies that longer window length provides more information for temporal characteristics, which aids in model improvement. Compared to TaKD$\ddagger$, TaKD$\dagger$ shows the better performance, which represents that utilizing more intermediate layers may act as a constraint on learning more diverse knowledge that would help students retain knowledge suited to its inherent characteristics \cite{zhang2019your, gou2021knowledge, wang2021distilling}.

\begin{table}[htb!]
\caption{Accuracy ($\%$) and $p$-value for various KD methods.}
\label{table:pvalue}
\centering
\begin{tabular}{c | c|c |c }
\hline
\centering
 & Method 1 & Method 2 & $p$-value\\
\hline
  \multirow{16}{*}{\rotatebox[origin=c]{90}{W100}} & \multirow{9}{*}{ \makecell{Student  \\ (6.458{\scriptsize$\pm$0.229})}}  & \makecell{KD\\ (6.554{\scriptsize$\pm$0.241})} & 0.034\\ 
   & & \makecell{SP\\ (6.606{\scriptsize$\pm$0.254})} & 0.012\\ 
   & & \makecell{KD+SP\\ (6.586{\scriptsize$\pm$0.253})} & 0.018\\ 
   & & \makecell{TaKD\\ (6.580{\scriptsize$\pm$0.252})} & 0.024 \\ 
   & & \makecell{TaKD$\dagger$\\ (6.394{\scriptsize$\pm$0.189})} & \textbf{0.023}\\ \cline{2-4}
  & \multirow{5}{*}{\makecell{KD\\ (6.554{\scriptsize$\pm$0.241})}}  & \makecell{SP\\ (6.606{\scriptsize$\pm$0.254})} & 0.098\\ 
   & & \makecell{KD+SP\\ (6.586{\scriptsize$\pm$0.253})} & 0.194\\ 
   & &  \makecell{TaKD$\dagger$\\ (6.394{\scriptsize$\pm$0.189})} & \textbf{0.079}\\ \hline
   
  \multirow{8}{*}{\rotatebox[origin=c]{90}{W200}} & \multirow{7}{*}{\makecell{KD\\ (6.284{\scriptsize$\pm$0.271})}}  & \makecell{SP\\ (6.236{\scriptsize$\pm$0.258})} & 0.125\\ 
   & & \makecell{KD+SP\\ (6.236{\scriptsize$\pm$0.258})} & 0.095\\ 
   & & \makecell{TaKD\\ (6.220{\scriptsize$\pm$0.225})} & 0.087 \\ 
   & & \makecell{TaKD$\dagger$\\ (6.201{\scriptsize$\pm$0.257})} & \textbf{0.022}\\ \cline{2-4}
   
\hline
\end{tabular}
\end{table}

We investigated the performance difference between a model learned from scratch (Student) and various KD methods. We conducted statistical analysis using $t$-tests with a confidence level of 90$\%$. Table \ref{table:pvalue} presents the RMSE ($\times$10$^{-2}$), standard deviation, and $p$-values for the models. For KD-based methods, Teacher3 was used as a teacher model. The $p$-values for pairs of Student and KD-based methods are less than 0.05, indicating a statistically significant difference between training from scratch and KD. For comparisons between  conventional KD and other methods, SP shows higher values compared to other methods. Additionally, as we expected, the $p$-value for KD and KD+SP is larger than 0.1 for W100 and approximately 0.1 for W200, suggesting that two methods are correlated.
Conversely, the $p$-values for KD with both TaKD and TaKD$\dagger$ are less than 0.1.
Based on these observations, we conclude that the proposed method demonstrates significantly different performance compared to other methods such as training from scratch and conventional KD, at a confidence level of 90$\%$.

\begin{figure}[htb!]
\includegraphics[scale=0.5] {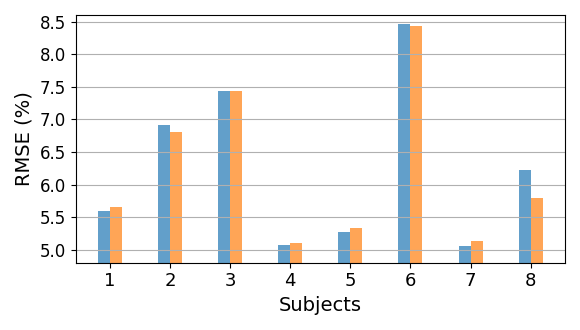} 
\centering
\caption{Results of Student (blue) and TaKD$\dagger$ (orange) for all subjects.}
\label{figure:all_Sbj}
\end{figure}

To show more details for each subject, we illustrated RMSE results of Student and TaKD$\dagger$ when Teacher1 is used. In Figure \ref{figure:all_Sbj}, learning from scratch (Student) shows higher errors than TaKD$\dagger$, where the averaged error is reported in Table \ref{table:grf_w100w200}. This represent that there is a variation and different statistical characteristics between subjects.

\subsection{Analysis on Autoencoder Methods}

To analyze the effectiveness of different autoencoders on our method, we compare AE \cite{kingma2013auto}, VAE \cite{fabius2015variational}, and WAE \cite{tolstikhin2018wasserstein} trained teachers in distillation. For WAE, the discriminator contains 5 linear and 4 ReLU layers and approximately 155M of total trainable parameters.

We trained a model with VAE, however, the results of RMSE is much higher than AE and WAE cases. In this particular study, due to the nature of the data, the variational nature of the VAE is not well suited for the task in hand \cite{sejnova2023benchmarking}. Introducing the VAE objective, specifically the latent distribution estimation, hinders the network performance \cite{daunhawerlimitations}. We
observe much higher RMSE for VAE than the other AE based
models.
To compare with more stable cases, we use AE and WAE trained teachers in KD. The results of teachers are explained in Table \ref{table:grf_wae}. In most of the cases, AE teachers perform better than WAE teachers.

\begin{table}[htb!]
\centering
\caption{Results of WAE teacher models for ground reaction force prediction. \textbf{Bold} denotes better results than the corresponding AE teacher.}

\begin{center}
\begin{tabular}{c |c c c |c c c}

\hline
\multirow{2}{*}{Method} & \multicolumn{3}{c|}{W200} & \multicolumn{3}{c}{W100} \\ \cline{2-7}



& RMSE & MAE & $r$ & RMSE & MAE & $r$ \\ \hline

\multirow{2}{*}{Teacher1} & \textbf{5.977} & 4.402 & \textbf{98.800} & 6.375 & 4.700 & \textbf{94.676} \\
& {\scriptsize$\pm$0.209} & {\scriptsize$\pm$0.141} & {\scriptsize$\pm$0.061} & {\scriptsize$\pm$0.219} &
{\scriptsize$\pm$0.150} & {\scriptsize$\pm$0.332} \\

\multirow{2}{*}{Teacher2} & 5.488 & 4.048 & 98.969 & \textbf{5.447} & 4.092 & 96.468 \\
& {\scriptsize$\pm$0.227} & {\scriptsize$\pm$0.163} & {\scriptsize$\pm$0.090} & {\scriptsize$\pm$0.256} &
{\scriptsize$\pm$0.200} & {\scriptsize$\pm$0.294} \\

\multirow{2}{*}{Teacher3} & 5.787 & 4.231 & 98.989 & 5.788 & 4.235 & 94.810 \\
& {\scriptsize$\pm$0.282} & {\scriptsize$\pm$0.192} & {\scriptsize$\pm$0.080} & {\scriptsize$\pm$0.266} &
{\scriptsize$\pm$0.173} & {\scriptsize$\pm$0.392} \\

\hline

\end{tabular}
\end{center}
\label{table:grf_wae}
\end{table}

\begin{figure}[htb!]
\includegraphics[scale=0.36] {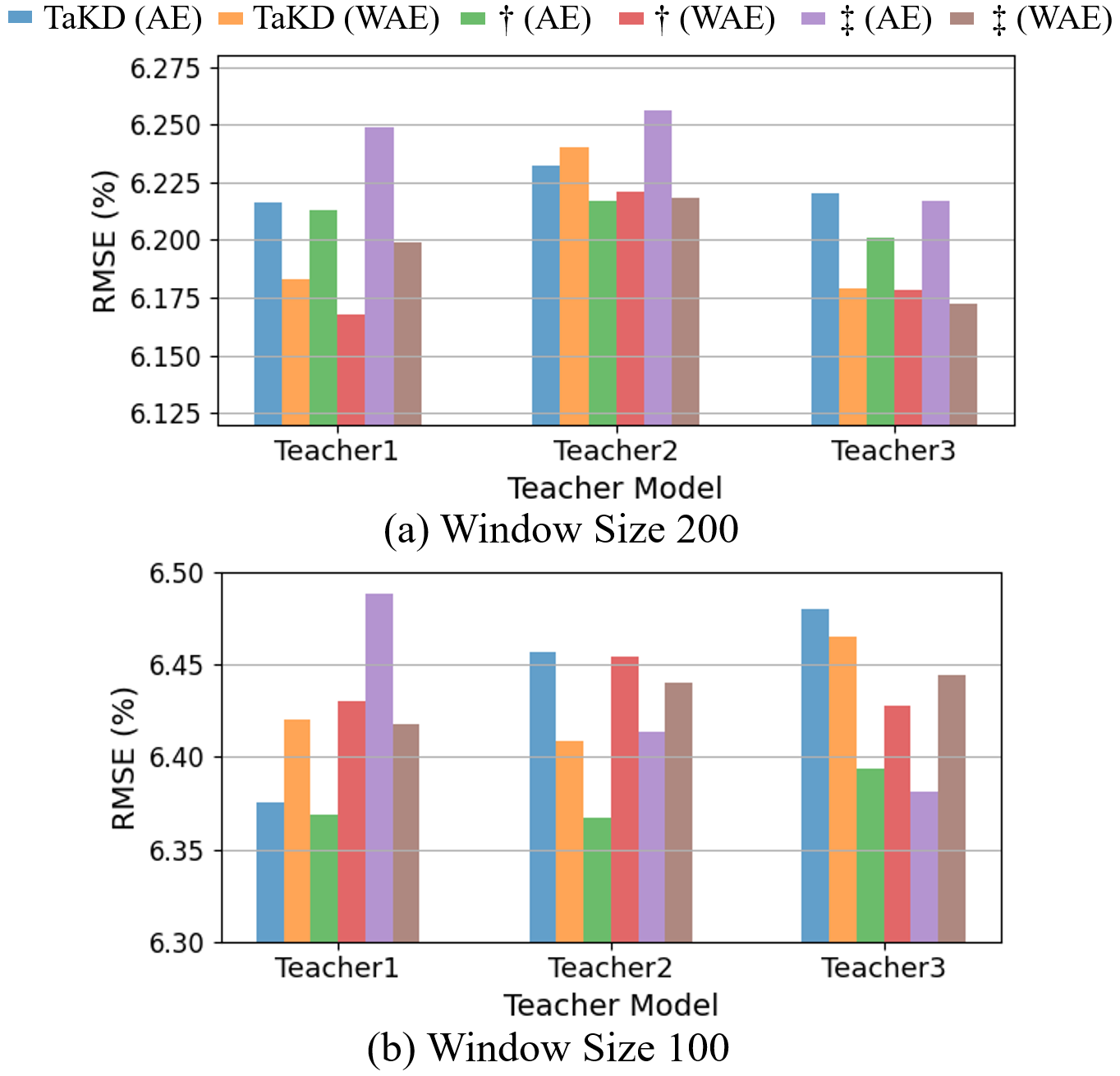} 
\centering
\caption{Results of students based on TaKD using teachers trained with AE and WAE for different window sizes.}
\label{figure:ae_wae_stu}
\end{figure}

As illustrated in Figure \ref{figure:ae_wae_stu}, when window size is 200, in most of the cases, WAE performs better than AE. This corroborates that better teacher does not guarantee superior student \cite{cho2019efficacy, jeon2022kd}. For window size 100, students from AE perform better, which implies that leveraging a regularizer of WAE is effective when sufficient information (e.g. patterns and characteristics of peaky and inflection points) is provided.

\subsection{Analysis on Learning Strategies}

\begin{figure*}[htb!]
\includegraphics[scale=0.52] {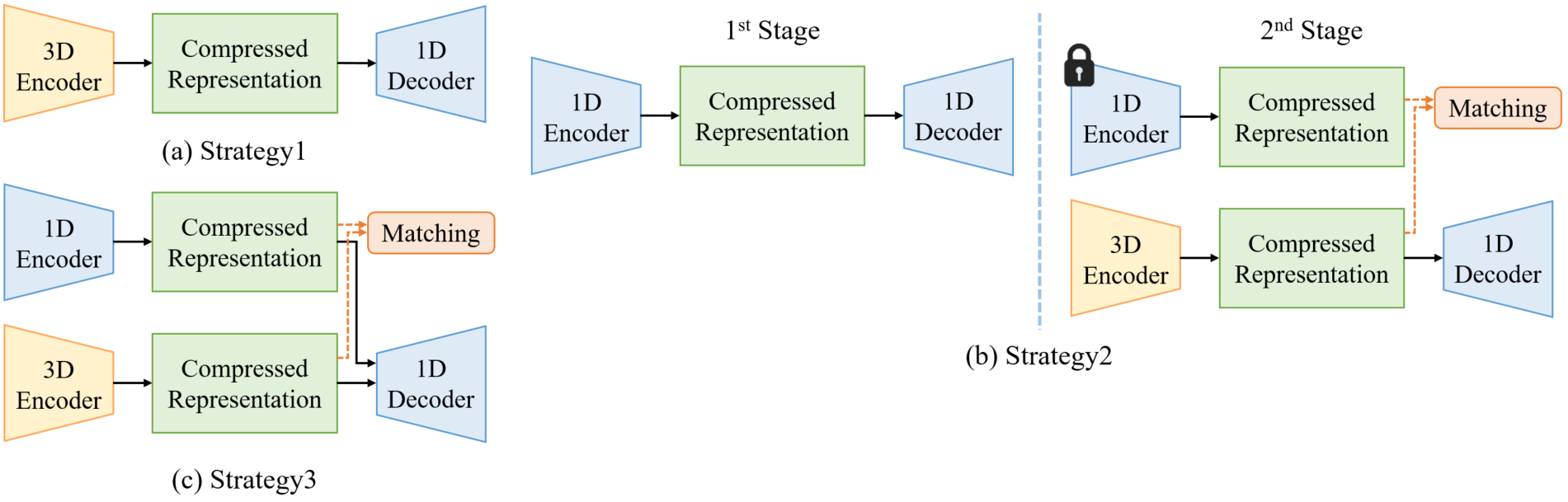} 
\centering
\caption{Illustration of strategies to train estimation models.}
\label{figure:strategy}
\end{figure*}

For analyzing crossmodal representations, many different strategies can be leveraged \cite{park2023cross, rajaram2023data}. We train student models with different strategies which are depicted in Figure \ref{figure:strategy}.
Strategy1 is a conventional way of AE, which is used for experiment of prior sections. Strategy2 includes two separated steps: 1) training 1D decoder with 1D encoder and 2) 3D encoder and 1D decoder are updated by leveraging the pretrained 1D encoder. Strategy3 is to leverage 3D and 1D encoders simultaneously. For Strategy2 and 3, to provide constraints to encourage the features from 3D and 1D encoders matched, cosine similarity based contrastive loss is used additionally. The weight parameter of 0.1 is used for the loss function. To evaluate on different strategies, we utilize TaKD$\dagger$, which showed best results in most of the cases for the previous sections. Teacher trained with Strategy1 is leveraged.
The results are explained in Table \ref{table:grf_strategy}. Strategy2 shows the better results among different strategies, which is effective for both AE and WAE, and outperforms baselines for overall cases reported in Table \ref{table:grf_w100w200}.
This represents that Strategy2 can be applied to our proposed method, TaKD, to improve performance effectively, which implies updating 1D decoder with 1D encoder is helpful for cross modality and estimation.


\begin{table}[htb!]
\centering
\renewcommand{\tabcolsep}{1.2mm} 
\caption{Comparison of predicted ground reaction force results (RMSE) with different strategies. \textbf{Bold} and \textcolor{red}{red} denote best and second-best results, respectively.}

\begin{center}
\scalebox{0.97}{
\begin{tabular}{c| c |c c c |c c c}

\hline
\multicolumn{2}{c|}{\multirow{2}{*}{Method}} & \multicolumn{3}{c|}{W200} & \multicolumn{3}{c}{W100} \\ \cline{3-8}



\multicolumn{2}{c|}{}& Teacher1 & Teacher2 & Teacher3 & Teacher1 & Teacher2 & Teacher3 \\ \hline


\multicolumn{2}{c|}{Student} & \multicolumn{3}{c|}{6.253{\scriptsize$\pm$0.259}} & \multicolumn{3}{c}{6.458{\scriptsize$\pm$0.229}} \\ \hline

\multirow{4}{*}{\rotatebox[origin=c]{90}{\scriptsize{Strategy1}}} &\multirow{2}{*}{AE} & 6.213 & 6.217 & 6.201 & 6.369 & \textbf{6.367} & 6.394 \\
& & {\scriptsize$\pm$0.252} & {\scriptsize$\pm$0.237} & {\scriptsize$\pm$0.257} & {\scriptsize$\pm$0.194} &
{\scriptsize$\pm$0.195} & {\scriptsize$\pm$0.189} \\

& \multirow{2}{*}{WAE} & 6.168 & 6.221 & 6.178 & 6.430 & 6.454 & \textbf{6.276} \\
& & {\scriptsize$\pm$0.215} & {\scriptsize$\pm$0.244} & {\scriptsize$\pm$0.263} & {\scriptsize$\pm$0.206} &
{\scriptsize$\pm$0.218} & {\scriptsize$\pm$0.200} \\


 \hline

\multirow{4}{*}{\rotatebox[origin=c]{90}{\scriptsize{Strategy2}}} & \multirow{2}{*}{AE} & \textcolor{red}{6.056} & \textcolor{red}{6.126} & \textcolor{red}{6.125} & \textcolor{red}{6.360} & 6.386 & \textcolor{red}{6.325} \\
& & {\scriptsize$\pm$0.225} & {\scriptsize$\pm$0.233} & {\scriptsize$\pm$0.240} & {\scriptsize$\pm$0.230} &
{\scriptsize$\pm$0.221} & {\scriptsize$\pm$0.215} \\

& \multirow{2}{*}{WAE} & \textbf{6.052} & \textbf{6.093} & 6.138 & \textbf{6.355} & \textcolor{red}{6.375} & 6.360 \\
& & {\scriptsize$\pm$0.233} & {\scriptsize$\pm$0.280} & {\scriptsize$\pm$0.240} & {\scriptsize$\pm$0.220} &
{\scriptsize$\pm$0.224} & {\scriptsize$\pm$0.217} \\

 \hline

\multirow{4}{*}{\rotatebox[origin=c]{90}{\scriptsize{Strategy3}}} & \multirow{2}{*}{AE} & 6.192 & 6.218 & \textbf{6.099} & 6.422 & 6.423 & 6.334 \\
& & {\scriptsize$\pm$0.239} & {\scriptsize$\pm$0.257} & {\scriptsize$\pm$0.230} & {\scriptsize$\pm$0.241} &
{\scriptsize$\pm$0.258} & {\scriptsize$\pm$0.223} \\

& \multirow{2}{*}{WAE} & 6.267 & 6.266 & 6.240 & 6.568 & 6.682 & 6.608 \\
& & {\scriptsize$\pm$0.208} & {\scriptsize$\pm$0.219} & {\scriptsize$\pm$0.224} & {\scriptsize$\pm$0.218} &
{\scriptsize$\pm$0.230} & {\scriptsize$\pm$0.211} \\

\hline

\end{tabular}
}
\end{center}
\label{table:grf_strategy}
\end{table}


\section{Ablations and Sensitivity Analysis} \label{sec:ablations}

In this section, we explore the effectiveness of layer selection in distillation, intermediate features, and hyperparameters on model sensitivity.

\subsection{Layer Selection for Distillation} \label{layer_selc}

In distillation, layer selection can affect to performance. We analyze with different combinations of layers. For the selected layers of encoder and decoder, similarity map within mini-batch is used for knowledge transfer. For the encoded representation transfer (Mid), temporal knowledge is additionally utilized. To show more gaps between methods, we evaluate various methods on smaller dataset consisting of approximately the half number of training and testing data with window size of 200. As explained in Table \ref{table:grf_inter1}, when outputs of E2 and D1 and Mid are used, the result shows best. Compared to using mid-level features (i.e. $\{$E1, D2$\}$), using high-level features (i.e. $\{$E2, D1$\}$) shows better results. Also, leveraging more intermediate layers generates better results.

\begin{table}[htb!]
\centering
\caption{Performance with various combinations of layers for the distillation. C3D teacher is used.}

\begin{center}
\begin{tabular}{c c c c c |c c c}

\hline
\multicolumn{2}{c}{Encoder} & & \multicolumn{2}{c|}{Decoder} & \multicolumn{3}{c}{Metrics} \\

E1 & E2 & Mid & D1 & D2 & RMSE & MAE & $r$ \\ \hline

\multirow{2}{*}{--} & \multirow{2}{*}{--} & \multirow{2}{*}{\checkmark} & \multirow{2}{*}{--} & \multirow{2}{*}{--} & 6.683 & 4.846 & 98.507 \\
& & & & & {\scriptsize$\pm$0.251} &
{\scriptsize$\pm$0.171} & {\scriptsize$\pm$0.131} \\ \hline

\multirow{2}{*}{--} & \multirow{2}{*}{\checkmark} & \multirow{2}{*}{--} & \multirow{2}{*}{\checkmark} &\multirow{2}{*}{--} & 6.678 & 4.835 & 98.520  \\
& & & & & {\scriptsize$\pm$0.250} &
{\scriptsize$\pm$0.173} & {\scriptsize$\pm$0.132} \\ \hline

\multirow{2}{*}{--} & \multirow{2}{*}{\checkmark} &\multirow{2}{*}{\checkmark} & \multirow{2}{*}{\checkmark} & \multirow{2}{*}{--} & 6.657 & 4.826 & 98.510 \\
& & & & & {\scriptsize$\pm$0.253} &
{\scriptsize$\pm$0.174} & {\scriptsize$\pm$0.134} \\ \hline

 \multirow{2}{*}{\checkmark} & \multirow{2}{*}{--} &\multirow{2}{*}{\checkmark} & \multirow{2}{*}{--} & \multirow{2}{*}{\checkmark} & 6.745 & 4.868 & 98.459 \\
& & & & & {\scriptsize$\pm$0.254} &
{\scriptsize$\pm$0.175} & {\scriptsize$\pm$0.135} \\ \hline

 \multirow{2}{*}{\checkmark} & \multirow{2}{*}{\checkmark} &\multirow{2}{*}{\checkmark} & \multirow{2}{*}{\checkmark} & \multirow{2}{*}{\checkmark} & \textbf{6.648} & \textbf{4.818} & \textbf{98.538} \\
& & & & & {\scriptsize$\pm$0.252} &
{\scriptsize$\pm$0.176} & {\scriptsize$\pm$0.130} \\ \hline

\end{tabular}
\end{center}
\label{table:grf_inter1}
\end{table}


\begin{table}[htb!]
\centering
\caption{Performance (RMSE) with various combinations of layers for the distillation.}

\begin{center}
\begin{tabular}{c c c c c |c c c}

\hline
\multicolumn{2}{c}{Encoder} & & \multicolumn{2}{c|}{Decoder} & \multicolumn{3}{c}{Teachers} \\

E1 & E2 & Mid & D1 & D2 & C3D & I3D & (2+1)D \\ \hline

\multirow{2}{*}{--} & \multirow{2}{*}{--} & bs & \multirow{2}{*}{--} & \multirow{2}{*}{--} &6.683 & 6.755 & 6.781 \\
& & +tp& & & {\scriptsize$\pm$0.251} &
{\scriptsize$\pm$0.263} & {\scriptsize$\pm$0.249} \\ \hline

\multirow{2}{*}{--} & \multirow{2}{*}{bs} & bs & \multirow{2}{*}{bs} &\multirow{2}{*}{--} &6.657 & 6.637 & 6.934  \\
& & +tp & & & {\scriptsize$\pm$0.253} &
{\scriptsize$\pm$0.251} & {\scriptsize$\pm$0.267} \\ \hline

\multirow{2}{*}{--} & bs & bs & bs & \multirow{2}{*}{--} & \textbf{6.639} & 6.689 & \textbf{6.770} \\
& +tp & +tp & +tp & & {\scriptsize$\pm$0.249} & {\scriptsize$\pm$0.258} & {\scriptsize$\pm$0.248} \\ \hline

 \multirow{2}{*}{--} & bs & bs & bs & \multirow{2}{*}{--} & 6.689 & \textbf{6.625} & 6.920 \\
& +ch & +ch & +ch & & {\scriptsize$\pm$0.253} &
{\scriptsize$\pm$0.250} & {\scriptsize$\pm$0.266} \\ \hline

\multirow{2}{*}{bs} & \multirow{2}{*}{bs} & bs & \multirow{2}{*}{bs} & \multirow{2}{*}{bs} & 6.648 & 6.667 & 6.962 \\
& & +tp & & & {\scriptsize$\pm$0.252} &
{\scriptsize$\pm$0.254} & {\scriptsize$\pm$0.262} \\ \hline

\multirow{2}{*}{bs} & bs & bs & bs & \multirow{2}{*}{bs} & 6.667 & 6.595 & 6.816 \\
& +tp & +tp & +tp & & {\scriptsize$\pm$0.252} &
{\scriptsize$\pm$0.242} & {\scriptsize$\pm$0.251} \\ \hline

\multirow{2}{*}{bs} & bs & bs & bs & \multirow{2}{*}{bs} & 6.692 & 6.674 & 6.929 \\
& +ch & +tp & +ch & & {\scriptsize$\pm$0.253} &
{\scriptsize$\pm$0.247} & {\scriptsize$\pm$0.257} \\ \hline
\end{tabular}
\end{center}
\label{table:grf_inter2}
\end{table}

We extend more evaluation with different types of teachers and knowledge maps. `bs', `tp', and `ch' denote using similarity map, temporal extent map, and channel similarity map, respectively, where the channel similarity map is represented with $Q = F \cdot F^{\top}; F \in \mathbb{R}^{c \times bthw}$. In Table \ref{table:grf_inter2}, the first, third, and sixth rows denote TaKD, TaKD$\dagger$ and TaKD$\ddagger$, respectively.
In overall, leveraging temporal knowledge in distillation shows effective results, compared to using other knowledge or combinations. In most of the cases, the third row (TaKD$\dagger$) shows better performance than others. This implies that using high-level features with temporal knowledge in distillation is effective. Also, using more intermediate layers can be acted as a constraint for performance improvement.

\subsection{Visualization of Intermediate Features}

\begin{figure*}[htb!]
\includegraphics[scale=0.54] {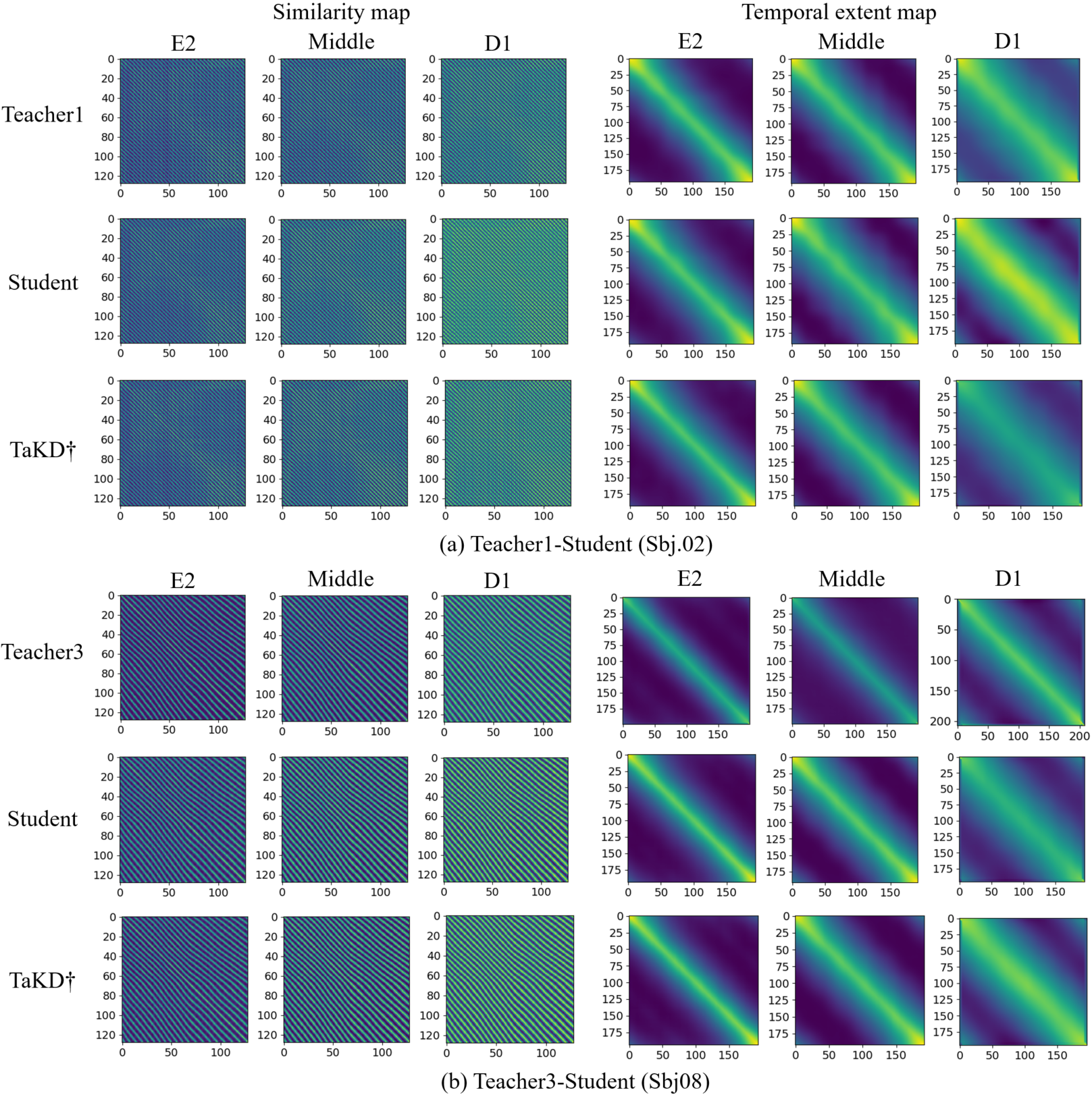} 
\centering
\caption{Activation maps representing similarity and temporal properties produced by the indicated layer of the network. Window size of samples is 200. The teachers are trained with AE.}
\label{figure:inter_sptp_map}
\end{figure*}

To provide richer knowledge in distillation, we use similarity and temporal extent maps. To understand their characteristics, we visualize the features from intermediate layers.
In Figure \ref{figure:inter_sptp_map}, we visualize the similarity and temporal maps from intermediate layers from teacher, student, and distilled student from TaKD$\dagger$ networks. For Teacher1-Student both combinations, the distilled student from TaKD$\dagger$ generates more similar patterns to Teacher1 compared to Student that is learned from scratch, which are intuitively shown for a decoder layer (D1). Specifically, in case (a), Student's D1 shows more brightness on diagonal points, however, TaKD$\dagger$ shows less brightness on the points than Student and is more similar to Teacher1. For case (b), Student shows less brightness on the points, whereas TaKD$\dagger$ shows brighter on the points and represents similar patterns to the one of Teacher3.
Also, these show that different subjects can possess different characteristics for activities and gait patterns, where the difference of periodical patterns is shown more with similarity map compared to temporal extent map. For temporal extent map, the brightness is different for subjects.
As explained in the previous sections and depicted in the figure, even though each subject has different statistical characteristics, our method, TaKD$\dagger$, can generate better estimation results compared to baselines and effectively transfer knowledge, which helps understanding temporal properties, to a student for improving performance.

\subsection{Analysis on Hyperparameters}
\begin{figure}[htb!]
\includegraphics[scale=0.31] {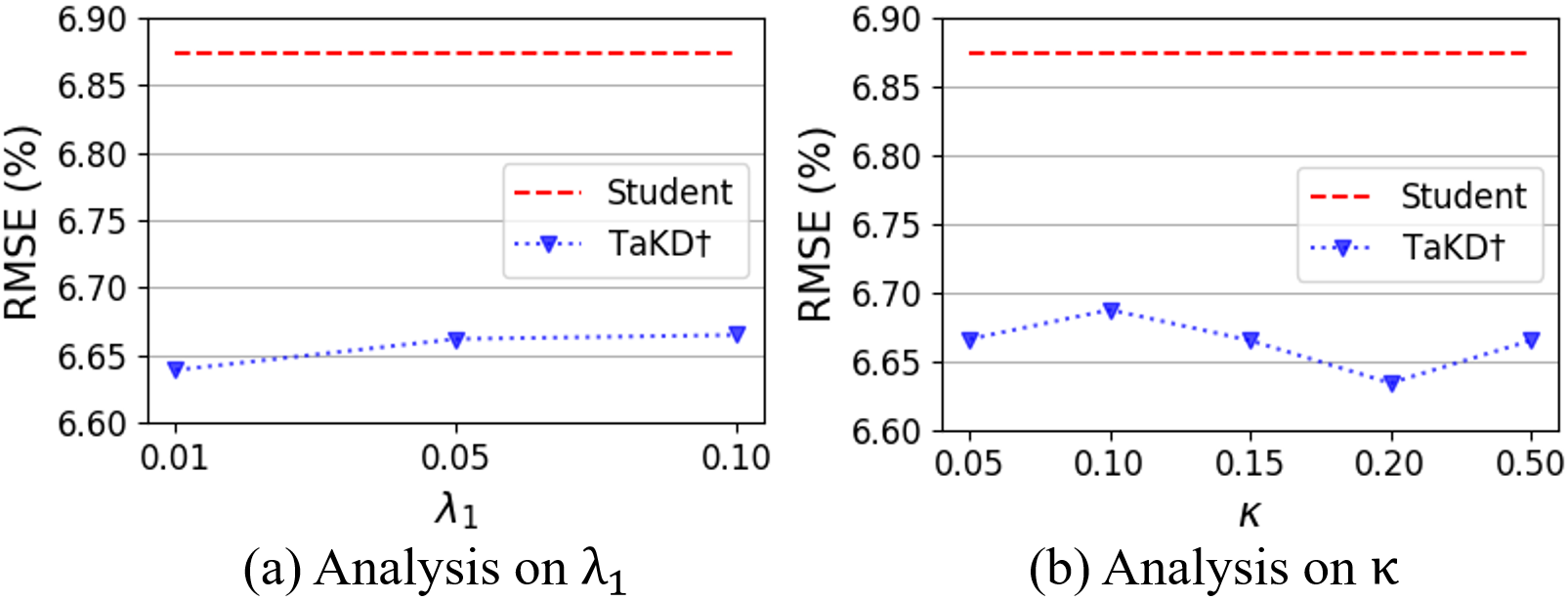} 
\centering
\caption{Results of TaKD$\dagger$ on various $\lambda_1$ and $\kappa$.}
\label{figure:ae_lambda_kappa}
\end{figure}

We utilize smaller $\lambda_1$, compared to $\lambda_2$ to control the effects between middle representation and and intermediate features. Smaller $\lambda_1$ implies providing less effects of middle representation. This is empirically selected in this paper, and the tendency is similar to prior studies \cite{jeong2022lightweight}. In general, compared to mid-level feature, reducing effects of high-level feature or middle representation is better in utilizing encoder and decoder networks for KD on regression or estimation problem.
As an ablation, we evaluate different combinations of $\lambda_1$ and $\lambda_2$ to explore more about their effectiveness and sensitivity. To show more gaps between methods, we evaluate various methods on smaller dataset consisting of approximately the half number of training and testing data with window size of 200. $\lambda_2$ is set as 0.1.
As shown in Figure \ref{figure:ae_lambda_kappa}(a), when $\lambda_1$ is lower than $\lambda_2$, the error is smaller. Thus, when smaller weights on middle representation than intermediate layers are used, TaKD$\dagger$ obtains better result. 

In Figure \ref{figure:ae_lambda_kappa}(b), we also evaluate the effects of different $\kappa$ to analyze the impact of temporal knowledge compared to similarity knowledge (1 is applied as a weight on similarity knowledge). When $\kappa$ is less than 0.5 (and larger than 0.1), the error is smaller. Additionally, as explained in the first to third row of Table \ref{table:grf_inter2}, leveraging temporal knowledge results in better distillation performance than either not using it or relying soley on the similarity map.
These imply that applying temporal knowledge generates a better student model and applying less the temporal knowledge than similarity shows better results.

\subsection{Model Reliability}

To investigate the model generalizability, we use expected calibration error (ECE) \cite{frank2015regression, naeini2015obtaining, guo2017calibration}. ECE is widely used to measure the miscalibration of a confidence measure, representing the reliability of the model. We adopt the metric to measure miscalibration between estimated GRF from a model and GRF data from a treadmill. T1 and T2 denote Teacher1 and Teacher2, respectively.
In Table \ref{table:ece}, ECE of various methods are explained. Compared to left foot data, the model has better reliability on right foot data estimation. This implies that the subjects participated in this experiments have body imbalance.
For averaged results, in T1, TaKD$\dagger$ shows best result. For T2, TaKD shows best. For Strategy2, WAE trained teacher can distill a better student in ECE performance, compared to using AE. In overall cases, leveraging Strategy1 is better in model reliability. This represent that model performing best reliability can differ from the one showing best RMSE results. Utilizing temporal extent aids in a better model distillation, not only for accuracy but also for reliability.

\begin{table}[htb!]
\centering
\renewcommand{\tabcolsep}{1.1mm} 
\caption{ECE ($\%$) for various methods on GRF estimation with window size of 200.}

\begin{center}
\begin{tabular}{c |c c c || c| c| c c c}

\hline

Method & Left & Right & Avg. & \multicolumn{2}{c|}{Method} & Left & Right & Avg.\\ \hline
Student & 2.011 & 1.654 & 1.832 & \multicolumn{2}{c|}{TaKD$\dagger$ (T1)} & 1.945 & \textbf{1.543} & 1.746  \\ \cline{1-4}

SP (T1) & 2.007 & 1.616 & 1.811 & \multicolumn{2}{c|}{TaKD$\dagger$ (T2)} & 2.165 & 1.813 & 1.989 \\  \cline{5-9}
SP (T2) & 1.909 & 1.654 & 1.781 & \multirow{4}{*}{\rotatebox[origin=c]{90}{\scriptsize{Strategy2}}} &  AE (T1) & 1.939 & 1.834 & 1.886 \\ \cline{1-4}
DIST (T1) & \textbf{1.810} & 1.682 & 1.747 & & AE (T2) & 1.965 & 1.792 & 1.879 \\
DIST (T2) & 2.140 & 2.002 & 2.071 & & WAE (T1) & 1.941 & 1.747 & 1.844 \\ \cline{1-4}
TaKD (T1) &  2.006 & 1.655 & 1.830 & & WAE (T2) & 1.940 & 1.830 & 1.885 \\ \cline{5-9}
TaKD (T2) & 1.864 & 1.626 & \textbf{1.745} & \multicolumn{5}{c}{}\\

\cline{1-4}

\end{tabular}
\end{center}
\label{table:ece}
\end{table}

\subsection{Visualization of Estimation Results} 
\begin{figure}[htb!]
\includegraphics[scale=0.39] {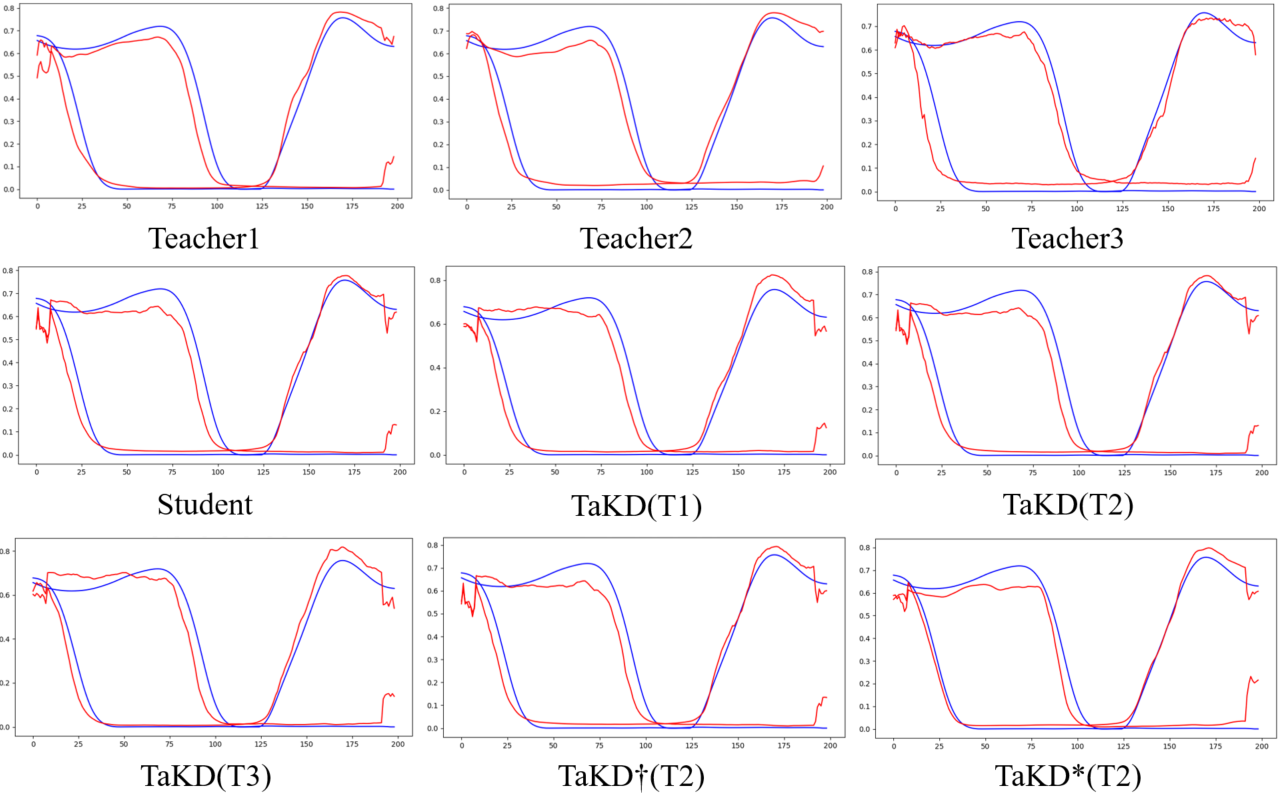} 
\centering
\caption{Illustration of the estimation results from various models for the SW condition. Blue and red colored lines denote GRF data (ground truth) and estimation result, respectively. * denotes a student distilled by TaKD with WAE teacher and Strategy2.}
\label{figure:recon_result1}
\end{figure}

\begin{figure}[htb!]
\includegraphics[scale=0.35] {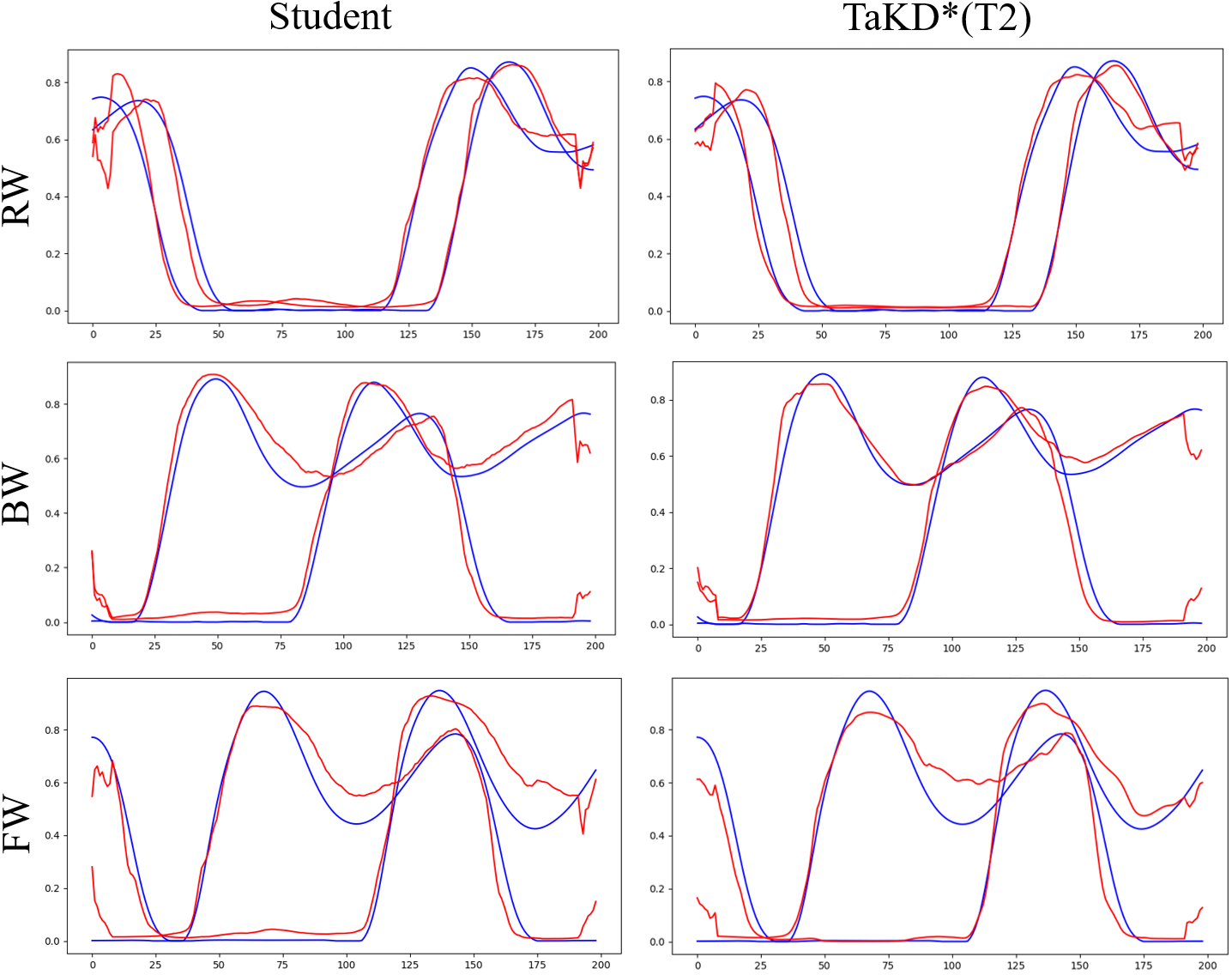} 
\centering
\caption{Illustration of the estimation results from various models for different walking speed conditions. Blue and red colored lines denote GRF data (ground truth) and estimation result, respectively. * denotes a student distilled by TaKD with WAE teacher and Strategy2.}
\label{figure:recon_result2}
\end{figure}

In Figure \ref{figure:recon_result1}, we visualize the estimation results from teacher and student models on a randomly chosen sample. T1, T2, and T3 denote Teacher1, Teacher2, and Teacher3, respectively. Since teachers have different architectural designs, their performance is different. Compared to Student that is learned from scratch, TaKD shows the less number of outlier points. Compared to TaKD$\dagger$, TaKD* shows better result, where TaKD* is by a student trained with WAE teacher and Strategy 2. In details, the difference of positions regarding to heel strike and toe-off between TaKD* and GRF (treadmill) data is smaller than the difference between Student and GRF data, where the difference plays a key role in estimation. 
Results across different walking conditions are described in Figure \ref{figure:recon_result2}. 
TaKD* performs better than Student in other movements as well. Especially at the inflection point, TaKD* shows better performance.


\section{Discussion} \label{sec:discussion}
As reported in the previous sections, TaKD showed robust performance in estimation and model reliability. In the estimation visualizations, the results near to starting and ending points have perturbations.
This is impacted by the size of the insole sensor, which was designed to fit a male with a foot size of US 9.5.
These perturbations can be resolved by using padded input data and then discarding the padded part to obtain the results from the model. Additionally, matching the positions of heel strike and toe-off can be improved using temporal warping functions. As presented in a previous study \cite{lohit2019temporal}, considering a function for time-series data, a temporal warping mechanism with sequential images can be developed further.




The GRF values obtained from the force plates during walking (measured in $N$) were normalized to each participant's weight to simplify computations in deep learning. When expressed as a percentage of bodyweight, a student model learned from scratch resulted in the following errors: RMSE of 7.07\% and 7.30\%, and MAE of 5.15\% and 5.35\% bodyweight for W200 and W100, respectively. 
Using our proposed method, a student model by TaKD$\dagger$ from Teacher2 for W100 achieved an RMSE of 7.19\% and MAE of 5.29\% bodyweight. Similarly, a student model based on TaKD$\dagger$ from Teacher3 for W200 yielded an RMSE of 7\% and MAE of 5.12\% bodyweight. Furthermore, with TAKD of WAE Teacher2 in Strategy2 (TaKD*), GRF estimation resulted in an RMSE of 6.84\% and MAE of 6.88\% bodyweight. These results demonstrate that our proposed methods and strategies distill effective and practically useful student models for GRF estimation.

The proposed method aims to solve issues related to different modalities, personal variability, and generating a small model for inference. Recently, several studies have focused on addressing problems in matching information by using different systems or conditioning in different environments. For example, one study utilized diverse sensors and systems to improve positioning performance \cite{seamless_systems}. Building on insights from the prior research, our study can be extended to features extraction for ensuring consistency or matching statistically dissimilar information during the stance phase for gait cycle, where the data originates from different systems.

To promote improved generalization, we extend this framework to include more subjects. Additionally, a broader range of activities, such as outdoor activities, walking on different inclined surfaces, and using different types of footwear, can be explored further.
Moreover, multi-modal sensors, such as inertial measurement units (IMUs), can be leveraged to enhance GRF estimation accuracy.

\section{Conclusion} \label{sec:conclusion}
In this paper, we introduced a robust framework-Time-aware Knowledge Distillation (TaKD), which leverages similarity and temporal extent properties to generate a lightweight model for GRF estimation. We evaluated the proposed method with different architectural networks, strategies, and combinations of teacher-student models, and reported results across diverse performance metrics.
Additionally, we designed a system and conducted tailored experiments, enabling the collection and analysis of diverse data types. We observed that leveraging feature relationships with temporal extent significantly enhances GRF estimation performance. Our findings provide insights for developing more advanced distillation methods for wearable devices and gait analysis.

The insole sensor used in this study has a fabric-like structure and is prone to warping while a person is wearing. Furthermore, the low resolution of the insole data provides very limited information for analysis, and the large variation in foot size among individuals affects the estimation performance in handling personal variability. Improved estimation results can be achieved if insole data was collected from sensors capable of extracting more precise and detailed information high resolution. Additionally, if the insole sensor was designed to account for factors such as gender, height, weight, and varying shoe sizes, more accurate results could be achieved. Extended studies using these sophisticated sensors may play a crucial role in the early diagnosis and treatment of diseases.

For future work, we aim to address the following: (1) Extending the framework to incorporate additional knowledge of predicted outputs, such as feature distributions and relationships, to provide stronger supervision in the distillation process, (2) Reducing the knowledge gap between student and teacher models by considering linear and non-linear distances caused by cross-modality characteristics, and (3) Extending this framework to include other health-related signals and  physiological parameter monitoring systems. In addition, our framework has significant potential for integration with other systems to match various types of information from different sensors or environments.

{\footnotesize
\bibliographystyle{IEEEtran}
\bibliography{main.bib}
}

\newpage

\vfill

\end{document}